\documentclass[printer]{aa}
\usepackage{amssymb}
\usepackage{amsmath}
\usepackage{natbib}
\bibpunct{(}{)}{;}{a}{}{,} 
\usepackage{graphicx}
\usepackage{txfonts}
\usepackage{color}
\usepackage{epstopdf}
\usepackage{wasysym}
\usepackage{multirow}
\usepackage{booktabs}

 \newcommand{\hv}[1]{\hat{\mathbf{#1}}}

\begin{document}

\title{Dust-enshrouded star near supermassive black hole: predictions for high-eccentricity passages near low-luminosity galactic nuclei}
\author{Michal Zaja\v{c}ek,\inst{1,2} Vladim\'{\i}r~Karas,\inst{1} \& Andreas Eckart\inst{3,4}}
\institute{Astronomical Institute, Academy of Sciences, Bo\v{c}n\'{\i}~II 1401, CZ-14100~Prague, Czech Republic  \and
Charles University in Prague, Faculty of Mathematics and Physics, V Hole\v{s}ovi\v{c}k\'ach 2, CZ-18000 Prague, Czech Republic \and
I. Physikalisches Institut der Universit\"at zu K\"oln, Z\"ulpicher Strasse 77, D-50937 K\"oln, Germany \and
Max-Planck-Institut f\"ur Radioastronomie (MPIfR), Auf dem H\"ugel 69, D-53121 Bonn, Germany}

\authorrunning{M.~Zaja\v{c}ek, V. Karas, \& A. Eckart}
\titlerunning{Dust-enshrouded star near supermassive black hole}
\date{Received 20 September  2013 / Accepted 18 March 2014}
\abstract{Supermassive black holes reside in cores of galaxies, where they are often surrounded by a~nuclear cluster and a~clumpy torus
of gas and dust. Mutual interactions can set some stars on a~plunging trajectory towards the black 
hole.}{We model the pericentre passage of a~dust-enshrouded star
during which the dusty envelope becomes stretched by tidal forces 
and is affected by the interaction with the surrounding medium. In particular, we explore under which conditions these encounters can lead to periods of enhanced accretion activity.}{We discuss different scenarios for such a dusty source. To this end, we employed a modification of the \texttt{Swift} integration 
package. Elements of the cloud were modelled as numerical particles that represent the dust component that interacts with the optically thin gaseous environment.}{We determine 
the fraction of the total mass of the dust component that is diverted from the original path during the passages through the 
pericentre at $\simeq10^3$ Schwarzschild radii and find that the main 
part of the dust ($\gtrsim90\%$ of its mass) is significantly
affected upon the first crossing. The fraction of mass captured at the second passage
generally decreases to very low values.}{As an example, we show
predictions for the dusty source evolution assuming the current orbital parameters of the G2 cloud (also known as Dusty S-Cluster Object, DSO) in our 
Galactic centre. Encounter of a core-less cloud with a supermassive black hole is, most likely, a non-repeating event: the cloud is destroyed. 
However, in the case of a dust-enshrouded star, part of the envelope survives the pericentre passage. We discuss an offset of $\lesssim0.3$ arcsec between the centre of mass of the diverted part and the star along the eccentric orbit. Finally, we examine an interesting possibility of a binary star
embedded within a~common wind envelope that becomes dispersed at the pericentre passage.}
\maketitle

\section{Introduction}
Most galaxies host supermassive black holes (SMBH; $10^6M_{\odot}\lesssim M_{\bullet} \lesssim 10^9M_{\odot}$) in their 
cores,  where these accrete gas and dust in the form of an accretion flow from their immediate neighbourhood 
\citep{1999agnc.book.....K,2012bhae.book.....M}. The example nearest to us is  the compact radio source Sgr~A*, which 
contains a black hole of mass $M_{\bullet}=4.4\times10^6M_{\odot}$ at distance $8.2$~kpc in the centre of the
Milky Way \citep{2005bhcm.book.....E,2007gsbh.book.....M,2010RvMP...82.3121G}. 

The character of accretion and the corresponding accretion rate vary greatly 
over different galaxy types. It appears that the availability of mass supply and the accretion mode that is established in the 
course of evolution of the system are the main agents that determine the power output and the spectral energy distribution of 
supermassive black holes \citep{2002apa..book.....F}. In several ways, the Galactic centre can serve as a paradigm for
low-luminosity nuclei.

Active galactic nuclei (AGN) and quasars host radiatively efficient types of disc accretion (i.e., the standard scheme of geometrically thin 
accretion discs, or slim discs; \citeauthor{1973A&A....24..337S} \citeyear{1973A&A....24..337S}; \citeauthor{1988ApJ...332..646A} 
\citeyear{1988ApJ...332..646A}) with accretion rates reaching and even exceeding the Eddington limit of 
$\dot{M}_{\rm{}Edd}\simeq L_{\rm{}Edd}/(0.1c^2)$, where
\begin{equation}
L_{\rm{}Edd}=\frac{4\pi GM_{\bullet}m_{\rm p}c}{\sigma_{\rm{}T}}\simeq1.3\times10^{44}\;\frac{M_\bullet}{10^6M_\odot}\quad[{\rm erg/s}],
\label{ledd}
\end{equation}
with $m_{\rm{}p}$ proton mass, $\sigma_{\rm{}T}$ Thomson cross-section.

Low-luminosity nuclei exhibit significantly lower accretion rates, $\dot{M}_{\bullet}\ll \dot{M}_{\rm{}Edd}$
\citep{2013PoS}. 
This can be explained as a combination of a diminishing supply of material falling onto the black hole and the
radiatively inefficient mode of accretion at certain stages.
In this context, the present state of the Galactic centre represents an extreme example of an inactive nucleus: 
$\dot{M_\bullet}\simeq10^{-8}M_{\odot}$ per year, which can be understood in terms of 
advection-dominated flow \citep{2008NewAR..51..733N}.\footnote{For
the supermassive black hole of Sgr~A* 
in the Galactic centre, the quiescent bolometric luminosity is $L_{\rm{}bol}=\eta\dot{M}_{\bullet}c^2\simeq10^{36}$ erg~s$^{-1}$. 
This corresponds to the dimensionless efficiency parameter for the conversion of accreted mass into radiation of about
$\eta\simeq10^{-3}$, although it can be as low as $10^{-5}$ at the present stage of the source.  The accretion outflow of Sgr A* is radiatively inefficient 
compared with predictions from the standard accretion disc theory, where $\eta\simeq 0.06$--$0.42$ is the predicted range.}

The temperature of the accreted material grows in the course of its infall in the gravitational field of the central black hole because the potential
energy is converted into heat and is only partially released in the form of emerging radiation \citep[e.g.][]{2002apa..book.....F,1999agnc.book.....K}. 
While at the distance of several tens to hundreds Schwarzschild radii 
($r_{\rm{}s}\equiv 2GM_{\bullet}/c^2\dot{=}2.95\times10^5\,M_{\bullet}/M_{\odot}$\,cm) the medium consists of ionised gas of the accretion 
disc and hot, diluted corona, farther out the temperature drops below the critical value for dust sublimation, 
$T_{\rm{}sub}\simeq 1.5\times10^3$~K
\citep{1987ApJ...320..537B,2005dusd.book.....K}. Therefore, at larger distances a clumpy torus can persist with a  
fraction of its mass in the form of dust \citep{1988ApJ...329..702K,2010A&A...523A..27H}.

An equilibrium can 
be reached through processes of dust sublimation (by strong irradiation from the central source and stars of the nuclear cluster), in competition with 
the replenishment of dust by stellar winds and the infall of clouds from the outer regions, where the circumnuclear 
torus is present \citep{1993ARA&A..31..473A,1995PASP..107..803U}. The co-evolution of gas and dust phases within clouds falling onto a supermassive 
black hole is relevant for our understanding of mass transport in the 
innermost regions of galactic nuclei. 

 Recently, an infrared-excess source named G2/DSO has been discovered \citep{2012Natur.481...51G} and subsequently detected in L- and K-bands \citep{2013ApJ...763...78G, 2013ApJ...773L..13P, 2013arXiv1311.2753E}. It may indeed be a manifestation of a common mechanism of material transport in low-luminosity nuclei. We analyse the scenario of an infrared-excess, dusty stellar source. As indicated in \citet{2014ApJ...783...31S}, the cloud component of the source is optically thin and diluted and not thick and dense. Therefore, it is valid to assume that the cloud component is mainly constituted by the gaseous wind driven by the radiation pressure of central star and the dust that is located and formed in such a wind. This is the reason that in the following analysis we assume the dust to be in contact only with stellar wind and the ambient atmosphere around Sgr A* through which G2/DSO travels.

 The adopted scenario is not necessarily only connected to this single event. It may be applied to other observed infrared-excess stellar sources that have been shown to move through the gaseous medium near the Galactic centre (e.g., \citeauthor{2005A&A...443..163M}, \citeyear{2005A&A...443..163M}, \citeauthor{2010A&A...521A..13M}, \citeyear{2010A&A...521A..13M}). Moreover, it may be relevant for modelling the environment in other low-luminosity active galactic nuclei.    

 In this paper we adopt a simplified (toy) model: dust grains are treated as numerical particles under the 
influence of gravity of SMBH ($M=M_\bullet$) and the embedded star ($M=M_\star$), or the components of a binary ($M_\star^{(1)}$, $M_\star^{(2)}$),
and the effect of an outflowing wind of gas. We focus on dust-enshrouded stars with different distributions of dust bound 
to the central star, and we explore the amount of material that is lost from the cloud to the black hole. Effects arise from the 
ambient pressure of a central wind, the wind pressure from the star, and a bow-shock forming at the interface of winds. The star moves at transonic speed near the pericentre.

 In this way we address the question whether and how dust particles embedded in the wind envelope are affected by close passages near the SMBH. For most model parameters, most of the dusty material is stripped from the envelope already on the first transit. For low accretion rates, the dust component can survive down to quite small radii, especially in regions shielded by obscuration. Furthermore, if the Field criterion \citep{1965ApJ...142..531F,2012MNRAS.424..728B} is fulfilled for the thermal stability of a two-temperature medium, the dust may co-exist with the hot medium at the same radius. The dust component by itself would be unimportant, but it contributes significantly to the radiation in NIR and tracing it helps to understand the observed emission. Similar treatment of dust dynamics is often employed in other astrophysical systems (mainly protoplanetary discs and stellar atmospheres, see e.g., \citeauthor{2011ApJ...734L..26V}, \citeyear{2011ApJ...734L..26V}).

 We model the encounters over a broad span of parameters. To present specific examples we use orbital parameters relevant for the Galactic centre G2/DSO infrared source \citep{2013ApJ...774...44G}, and we also attempt to distinguish among different outcomes of the passages through the pericentre \citep{2013A&A...551A..18E,2013ApJ...773L..13P}.

\begin{figure*}[tbh]
\centering
 \includegraphics[width=0.32\textwidth]{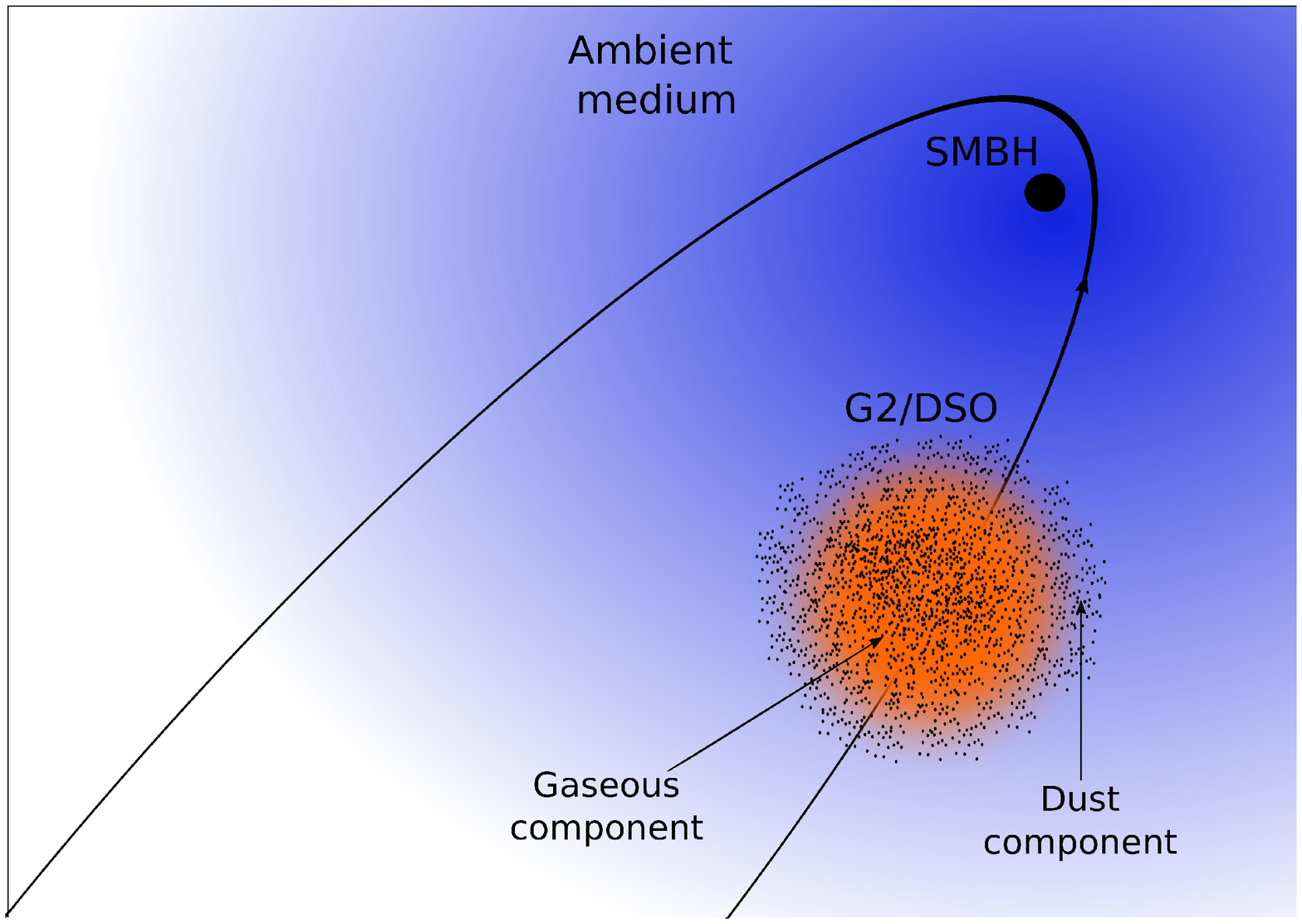}  
 \includegraphics[width=0.32\textwidth]{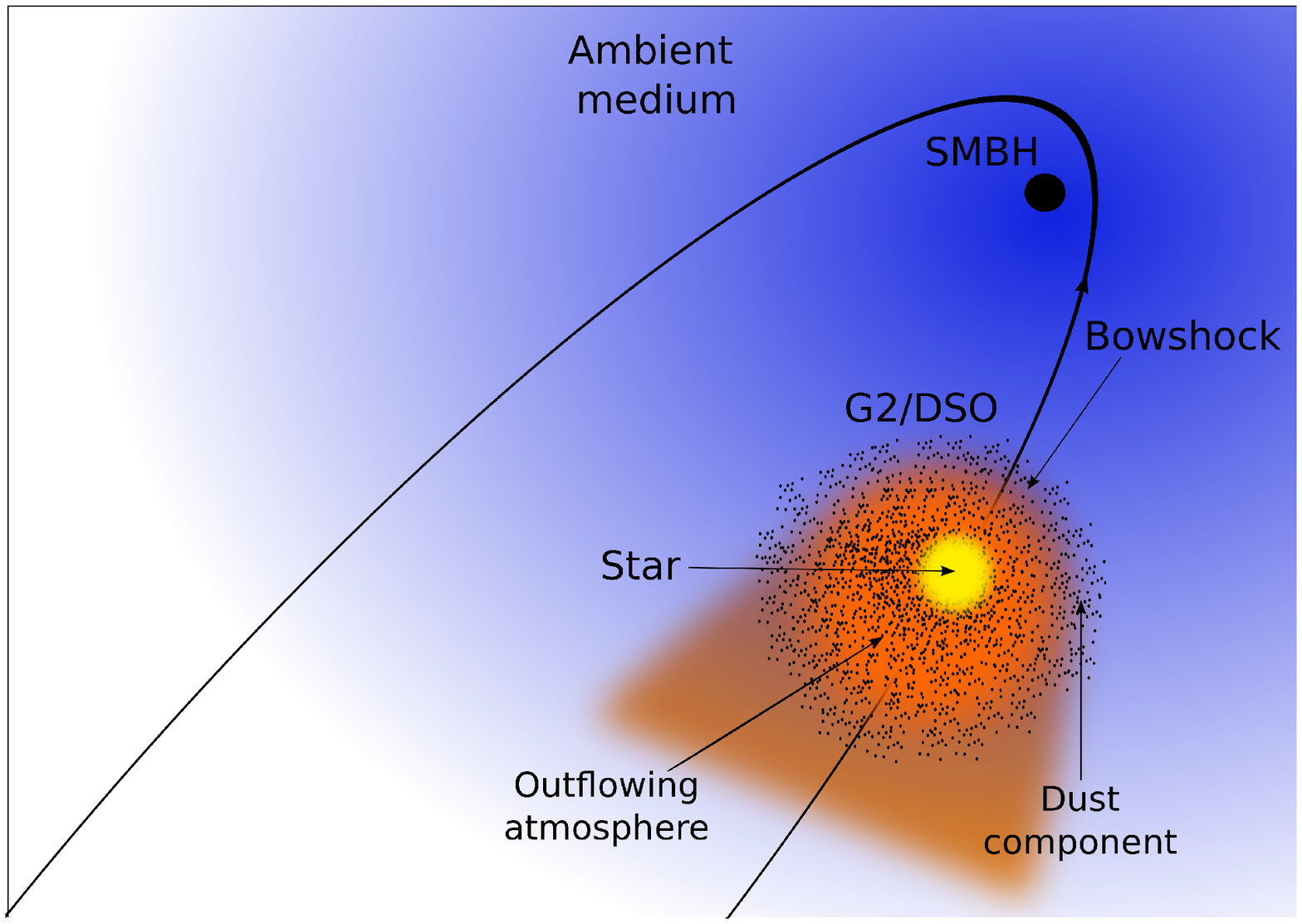}  
 \includegraphics[width=0.32\textwidth]{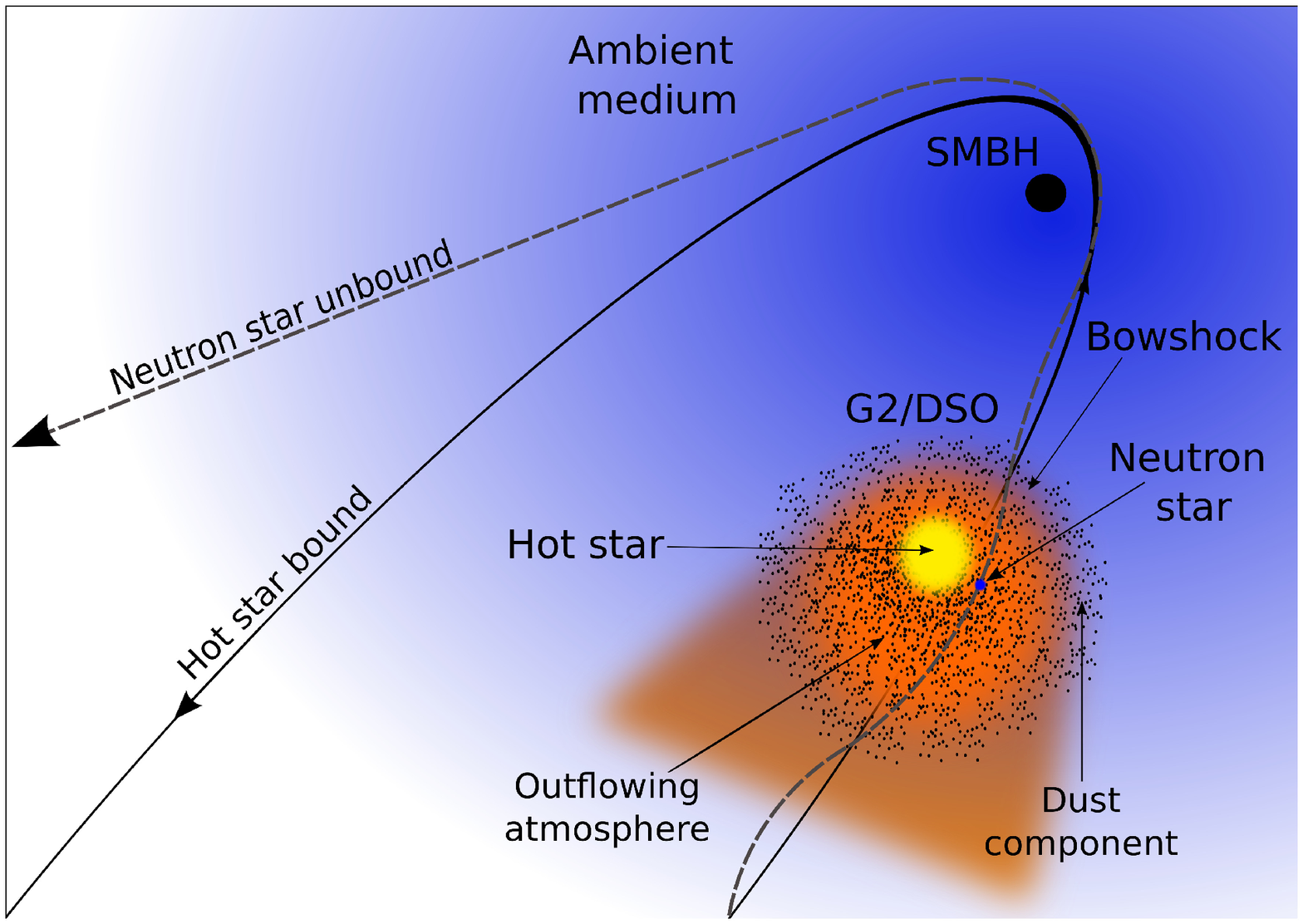}  
\caption{Three variants of the model setup for which the predictions are qualitatively different especially at the post-pericentre phase. 
The model ingredients include the central supermassive black hole (black circle), an infalling cloud made of gas and dust (red),
an embedded star (yellow), and a hot diluted flow (blue).  
We consider the pericentre at about $10^3$ Schwarzschild radii, so that the star is not expected to be tidally disrupted. 
However, the gaseous/dusty envelope is affected significantly. 
\textit{Left panel:} a core-less cloud on an eccentric trajectory, interacting with the diluted ambient medium near the SMBH.
\textit{Middle panel:} the cloud enshrouds an embedded star. A radial wind of the outflowing gaseous atmosphere occurs 
and a bow-shock forms ahead of the stellar body. \textit{Right panel:} binary system enclosed by a common envelope that becomes largely dispersed
at the first pericentre passage. At the same time, the three-body interaction with the central SMBH causes the binary components 
to separate from the nominal trajectory.}
\label{illustration} 
\end{figure*}

Furthermore, we point out to the possibility that the stellar core may actually consist of two components of a binary star. This idea is suggested by models of the origin of S-stars in the Galactic centre as a product of three-body interaction during the pericentre passage of a binary star on a highly eccentric trajectory \citep{2003ApJ...592..935G,2007ApJ...656..709P}. Although the presence of a stellar core and its putative binary nature within the G2 cloud are on a purely hypothetical level, this scenario can connect, in a natural way, two apparently different aspects: the high eccentricity
of the plunging trajectory, and the origin of the population of stars near the supermassive black hole. If there is indeed a star enshrouded by a dusty atmosphere, it was proposed that high eccentricity can be achieved by the Kozai mechanism \citep{2005A&A...433..405S} or by resonant relaxation \citep{2006ApJ...645.1152H}. 

The geometrical setup and the main ingredients of our model are illustrated in figure~\ref{illustration}.
Three different flavours of the basic scenario were considered: a core-less cloud infalling onto the
SMBH, a star embedded within the dusty envelope, and an embedded binary that becomes 
disrupted near the SMBH. We focus on the latter two scenarios.
Our simplified approach is complementary to purely hydrodynamical situations that neglect the dust component  
\citep{2012ApJ...759..132A,2012ApJ...750...58B,2013ApJ...776...13B,2014ApJ...783...31S}, which is consistent with an optically thick, dense medium where the dust is dragged along with the gas. However, for optically thin atmospheres, dust dynamics needs to be treated separately. 

In the following analysis, we do not treat Br$\gamma$ production or the radiation processes in the bow-shock region \citep{2013MNRAS.433.2165S}.
We do see, however, that the dusty envelope is stretched by the gravitational and drag forces (depending on the initial distribution of particles in phase space and the parameters of the wind outflow), which leads to the gradual offset between the dust component and the stellar core. We note that observationally any difference between the cloud location in L-band with respect to the location
of K (Br$\gamma$) in the orbit is most likely due
to uncertainties in the determination of the orbital positions; it may be heavily affected by different systematics in the two bands.

 The paper is organized as follows: in sect. \ref{sec1} we set
up the model and describe the numerical procedure to explore the mutual 
interaction between the star and its environment. We discuss the dependence of the dust temperature on the distance 
and the luminosity of the central source. Then we consider the effect of the star enshrouded by an initially spherical dusty envelope and a remnant disc. In sect. \ref{sec2}, we present the results of 
the simulations including the wind blowing from the centre and the effect of the bow-shock region. We  
compare the difference between a disc-like Keplerian distribution and a Gaussian distribution of particles in the phase space. Finally, we 
determine the fraction of dust mass affected at subsequent encounters, and we show the offset that develops gradually between the centre of mass
of the cloud and the nominal position of the star in the orbit. In sect. \ref{discussion} we summarize and discuss our results, and we 
conclude in sect. \ref{conclusions}.

\section{Model setup, numerical scheme, and tests}
\label{sec1}
\subsection{Clouds infalling onto SMBH - test runs}
Accretion tori are not smooth, instead, they often form individual clouds with a mixed composition of gas and dust phases 
\citep[e.g.,][]{2005ARA&A..43..337C,2005pacp.book.....V}. The patchy 
structure evolves by collisions, and some of the clumps can be set on a highly eccentric trajectory with the pericentre close to the central 
black hole. Interaction with the surrounding environment leads to the drift and gradual separation of different constituents of the 
cloud and deformation of its shape -- dust and gas components of the cloud move with respect to each other, and with respect to the star 
(if embedded in the cloud core). 

A bow shock develops around an embedded star and different species are transported
across discontinuities in a different manner \citep[smaller grains tend to be bound to the gas component, while the large grains are capable of 
penetrating into the interstellar medium; cf.][]{1992ApJ...397..644M,2011ApJ...734L..26V,2012ASPC..459...73K}. If large
portions of the medium are neutral, viscous forces play an important role, while 
in a fully ionised plasma the efficient mechanism of dust grain charging operates. Then the main parameter that defines the coupling between dust and gas is the Coulomb coupling
parameter, which is the ratio of the Coulomb potential energy of the particle interaction to the kinetic energy of the thermal motion.
Depending on the system parameters, both weak coupling and strong coupling have been observed in dusty plasmas.

The grain size is a dominant factor for the forces acting on the gas/dust mixture. The mutual coupling between these two 
components of the complex plasma (including the viscous forces) is thought to just increase 
the effective dust grain mass by the amount of gas that is dragged
along in  the coupling. The particle mass is not relevant within the strong gravitational field of the SMBH and/or the star,
instead, it may only be relevant for the wind force acting on the dust. For a gas-to-dust ratio of 100:1, the effective 
increase of the grain mass load by two orders of magnitude would then correspond to the grain size variation by factor of 
about 5. This is well contained in the factor of a hundred in the grain sizes that we considered here.
The relation between size and mass is uncertain because grains are most likely fluffy and not solid. 
Moreover, other factors are also connected with the grain size, for example, the typical electric charge
that can develop by the interaction with the surrounding plasma and by photoionisation.

An infalling cloud passes through the external environment with physical properties spanning a wide range of values. We modelled the 
passages of stars with dusty envelopes through the pericentre by numerically integrating the trajectories of star(s) and that 
of dust particles that represent one of elementary constituents of the cloud. These are followed in the gravitational field of the SMBH, taking into account
the hydrodynamical interaction with the diluted ambient wind. Gravitational effects need to be taken into account, but
nonetheless, the Newtonian description is adequate for the gravitational field
because we considered the motion with the pericentre at $\simeq10^3r_{\rm{s}}$, so not in the immediate vicinity of the black hole
horizon, where the relativistic effects on the orbit evolution become important.
The variety of factors listed above underline a potentially important influence
of the grain properties for the gas/dust coupling and hence support our simulation approach. 

\begin{figure}[tbh]
\includegraphics[width=0.5\textwidth]{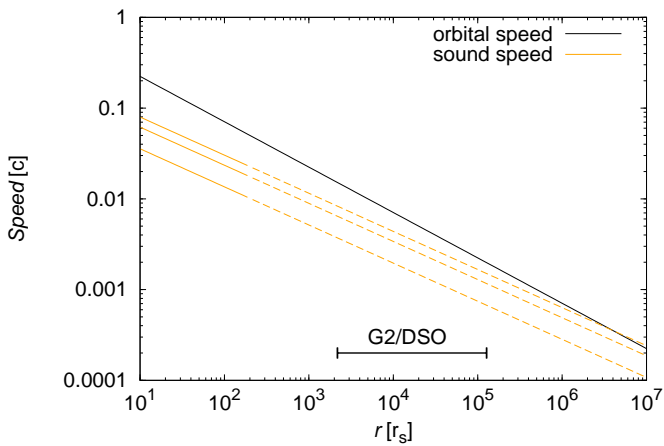}
\caption{Comparing the orbital (Keplerian) velocity with the sound speed at corresponding radius
from the central black hole (units of Schwarzschild radii). The range of G2/DSO  is 
labelled. On the vertical axis, the unit of speed is the speed of light, $c$.}
\label{img_velocity}
\end{figure}

\begin{figure}[tbh]
\centering
\includegraphics[width=0.5\textwidth]{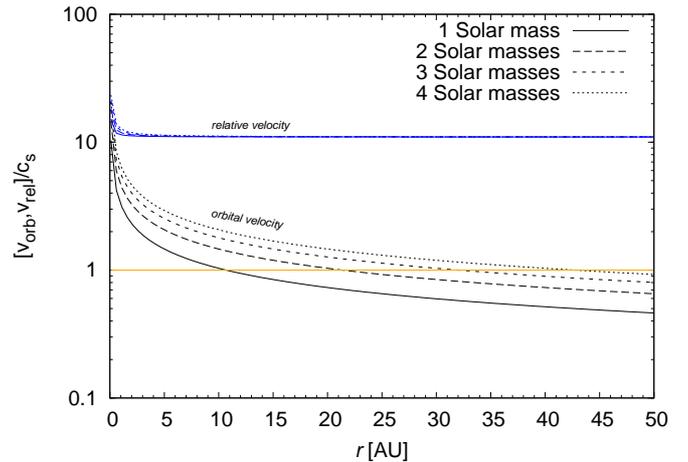}
\caption{Ratio of orbital velocities (black lines) and relative velocities (blue lines) to the speed of sound as a function of distance from the star. Relative velocities are typically higher than the sound speed by about one order of magnitude for wind velocities of the order of $100\,\rm{km\,s^{-1}}$.}
\label{img_orbvel_cs}
\end{figure} 

\begin{figure}[tbh]
\centering
\includegraphics[width=0.5\textwidth]{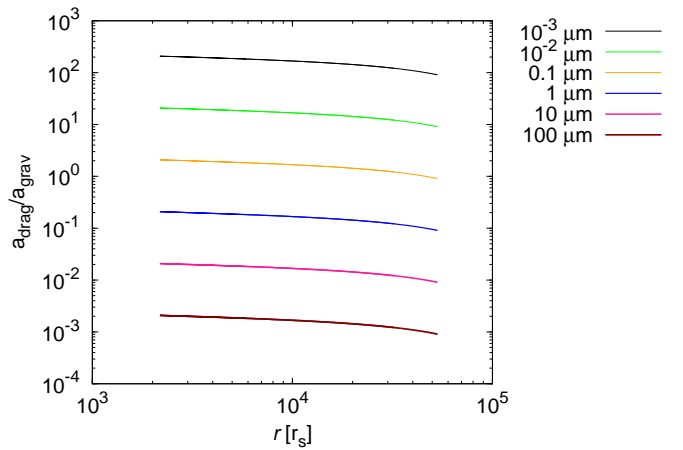}
\caption{Ratio of the hydrodynamical drag to the gravitational 
acceleration across the range of radii.  The two effects can be
of comparable magnitude and, therefore, both need to be taken into account in the model of the orbit evolution.}
\label{img_comparison_acceleration}
\end{figure} 

\begin{figure}[tbh]
\centering
\includegraphics[width=0.5\textwidth]{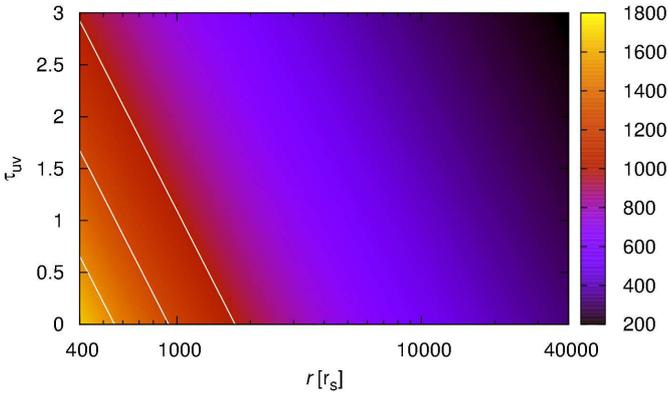}
\caption{Colour-coded profile of temperature $T\equiv T(r,\tau_{uv})$ of dust for the central 
source of luminosity $10^5\,L_{\odot}$ near the central SMBH. From right to left, 
the three lines are isotherms corresponding to $T=1000$~K, $1250$~K, and $1500$~K 
(dust sublimation temperature).}
\label{img_tempcomp}
\end{figure} 

We employed the integration package \texttt{Swift} \citep{1994Icar..108...18L},
which has been designed to evolve 
a set of mutually gravitationally interacting bodies together with a group of test particles that are influenced by gravity of 
the massive bodies, but do not affect each other (or the massive bodies). In the numerical scheme we included the effect of radial wind outflow and that 
of ambient gaseous wind and employed a sufficiently accurate Bulirsch-Stoer integrator with an adjustable time-step. 

The adopted procedure allowed us to follow a large number of numerical particles 
in the gravitational field, namely, a superposed gravitational field of the
central SMBH and the embedded star. The gravity of the black hole plays a role in the motion of parcels because their
mass is relatively large (like dust grains compared to gas atoms), but we modified the integration routine to also include the hydrodynamic 
drag that acts through the interaction of particles with the gaseous ISM as well as an outflowing stellar wind. 
The dust grain mass is a parameter that allowed us to
study different cases, including a toy model where the grains represent super-particles of mass exceeding that of realistic 
cosmic dust.

\subsection{Non-gravitational forces acting on dust}
\label{sec_hydro_drag}
 Although the stellar atmosphere consists largely of outflowing gas, here we concentrated mainly on 
the dust component. Both species interact with each other, but they can separate to a certain extent during the evolution.
Observationally, the gas is expected to be revealed by Br$\gamma$ spectral line emission, while the dust grains will contribute to the 
thermal continuum and  can be detected by polarisation. This
provides information about in situ conditions of the surrounding environment, but the predictions are currently
uncertain due to incomplete knowledge about the nature and composition
of G2/DSO.

In a galactic nucleus all objects interact with the hot plasma that is supplied by stellar winds, in particular, of hot, massive OB 
stars. The effects of this interaction depend on temperature 
and density profiles of the ISM. For the purpose of this work, we used semi-analytical relations for electron density 
$n_{\rm{e}}(r)$ and electron temperature $T_{\rm{e}}(r)$ radial profiles based on the models 
of radiatively inefficient accretion flows \citep{2006ApJ...636L.109B,2011ApJ...738...38B},
\begin{equation}
  n_{\rm{e}}(r)  = n_{\rm{e}}^{0}\left(\frac{r}{r_{\rm{s}}} \right)^{-1.1} \label{density_profile},\quad
  T_{\rm{e}}(r)  = T_{\rm{e}}^{0}\left(\frac{r}{r_{\rm{s}}} \right)^{-0.84} 
.\end{equation}
In these dependencies, we considered quantities 
$n_{\rm{e}}$ and $T_{\rm{e}}$ to adopt averaged values,  
$n_{\rm{e}}^{0}=3.5 \times 10^7\,\rm{cm^{-3}}$ and $T_{\rm{e}}^{0}=9.5 \times 10^{10}\,\rm{K}$. 
Electrons are decoupled from ions at small distances from the SMBH; ion temperatures are $\sim$ 1 to 5 times higher than 
electron temperatures resulting from MHD simulations \citep[e.g.,][]{2010ApJ...717.1092D}. The density profile 
\eqref{density_profile} is used in eq.\ \eqref{eq_drag_term} to estimate the ambient density 
$\rho_{\rm{a}}=m_{\rm{H}}\, n_{\rm{H}}$ (with $n_{\rm{H}} \approx n_{\rm{e}}$). 

\begin{figure*}[tbh]
\centering
 \includegraphics[width=0.49\textwidth]{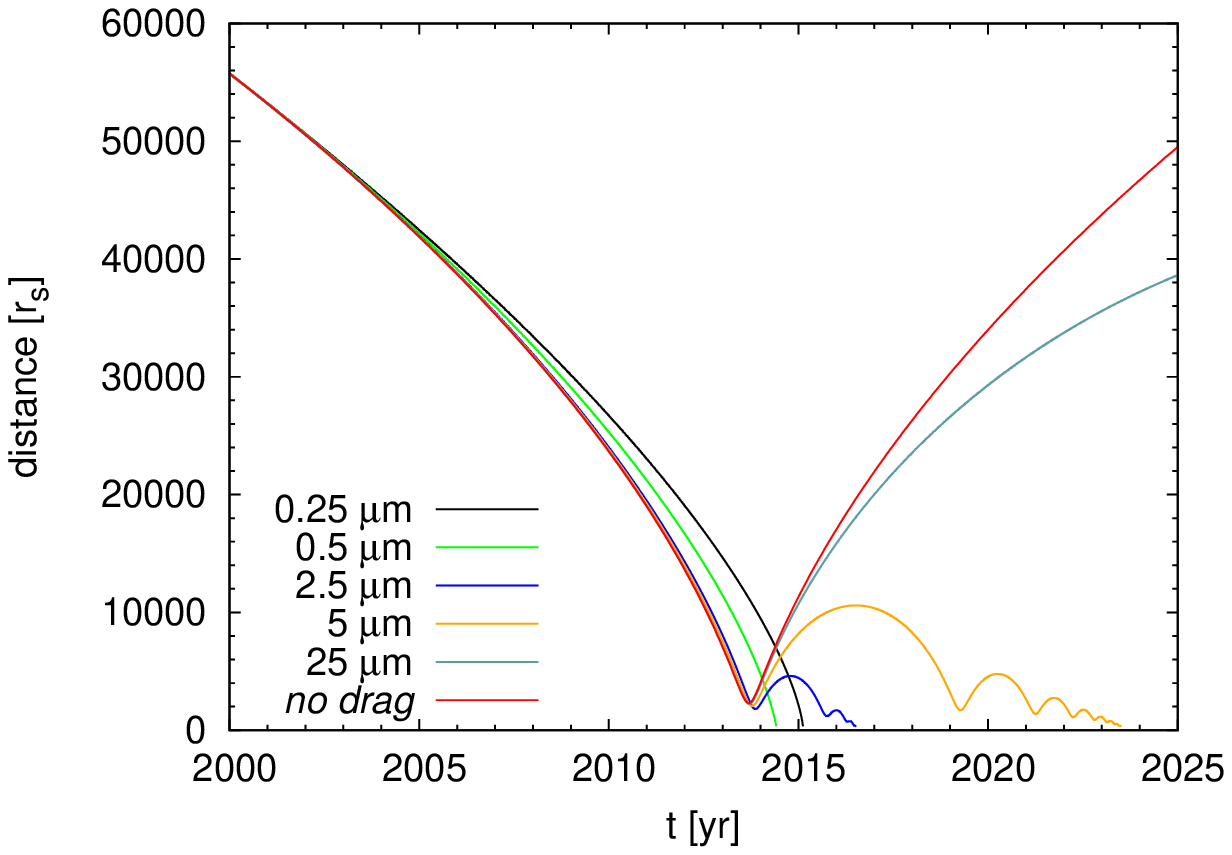}  
 \includegraphics[width=0.49\textwidth]{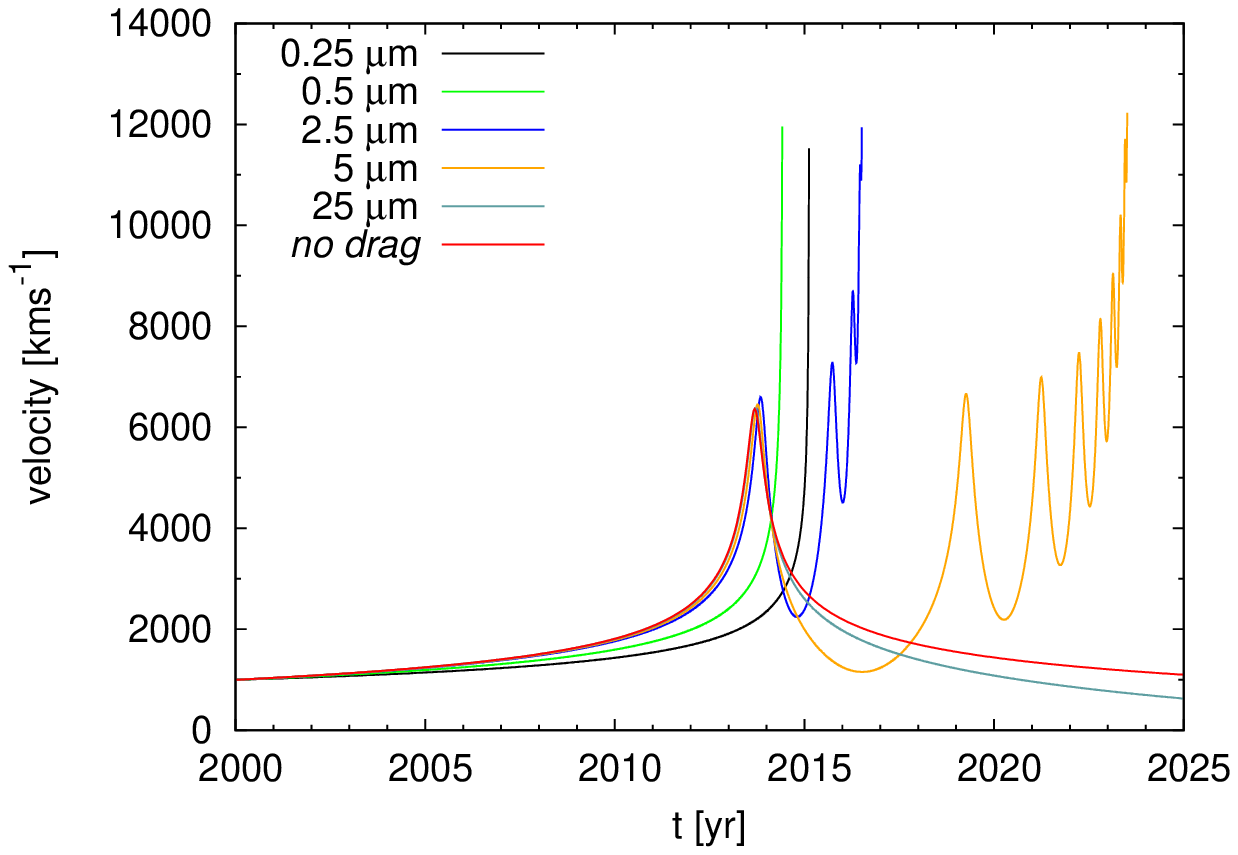}
\caption{
Single-particle approximation.
\textit{Left panel:} the temporal evolution of 3D distance (in Schwarzschild radii); the orbital parameters are set consistent 
with the G2/DSO in Galactic centre. Different particle (dust grain) sizes are considered, taking into account the effect of gravity 
and the hydrodynamical drag exerted by the ambient medium.  
\textit{Right panel:} the corresponding velocity as a function of time. Hydrodynamical drag acts more efficiently on particles of
larger size, causing their rapid in-spiralling and accretion onto the central black hole.}
\label{img_dbh_v3d} 
\end{figure*}

\begin{figure*}[tbh]
\centering
  \includegraphics[width=0.49\textwidth]{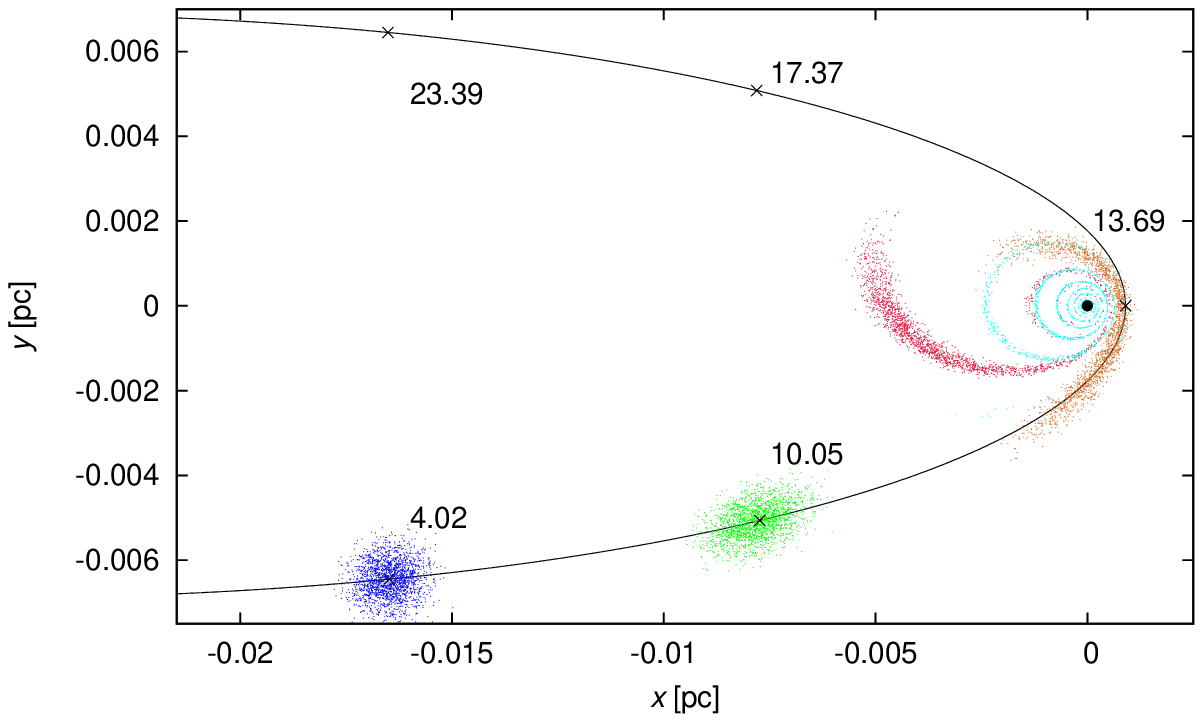} 
  \includegraphics[width=0.49\textwidth]{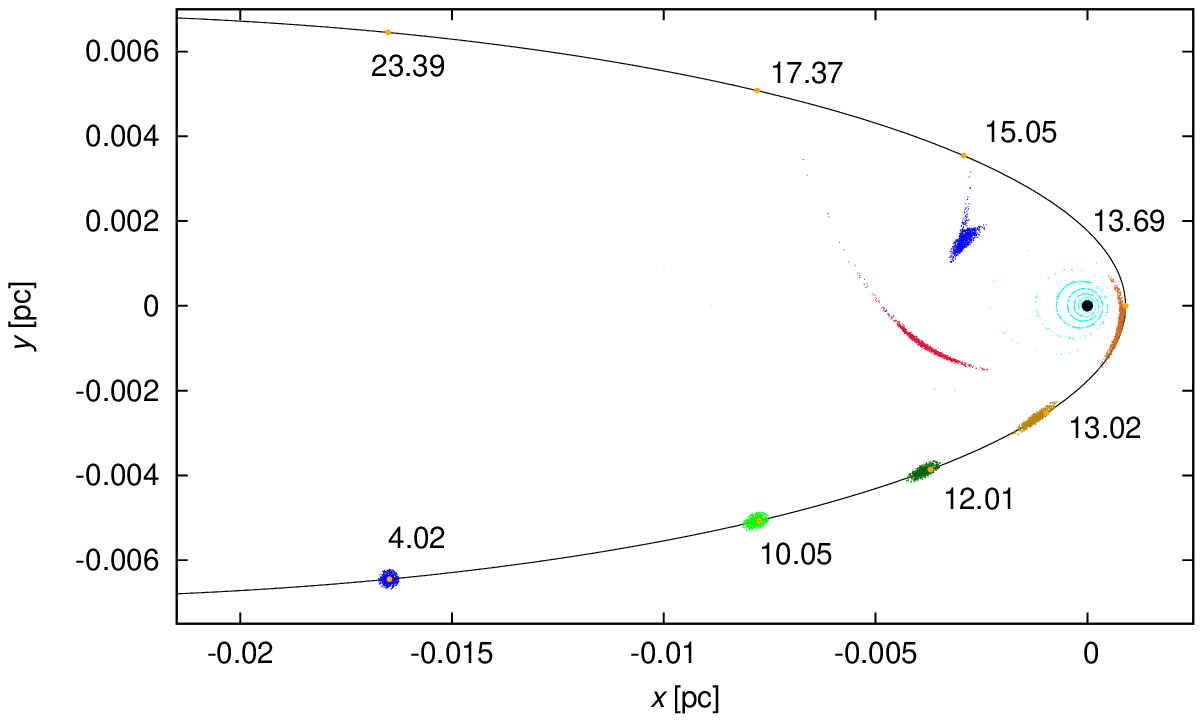} 
  \includegraphics[width=0.49\textwidth]{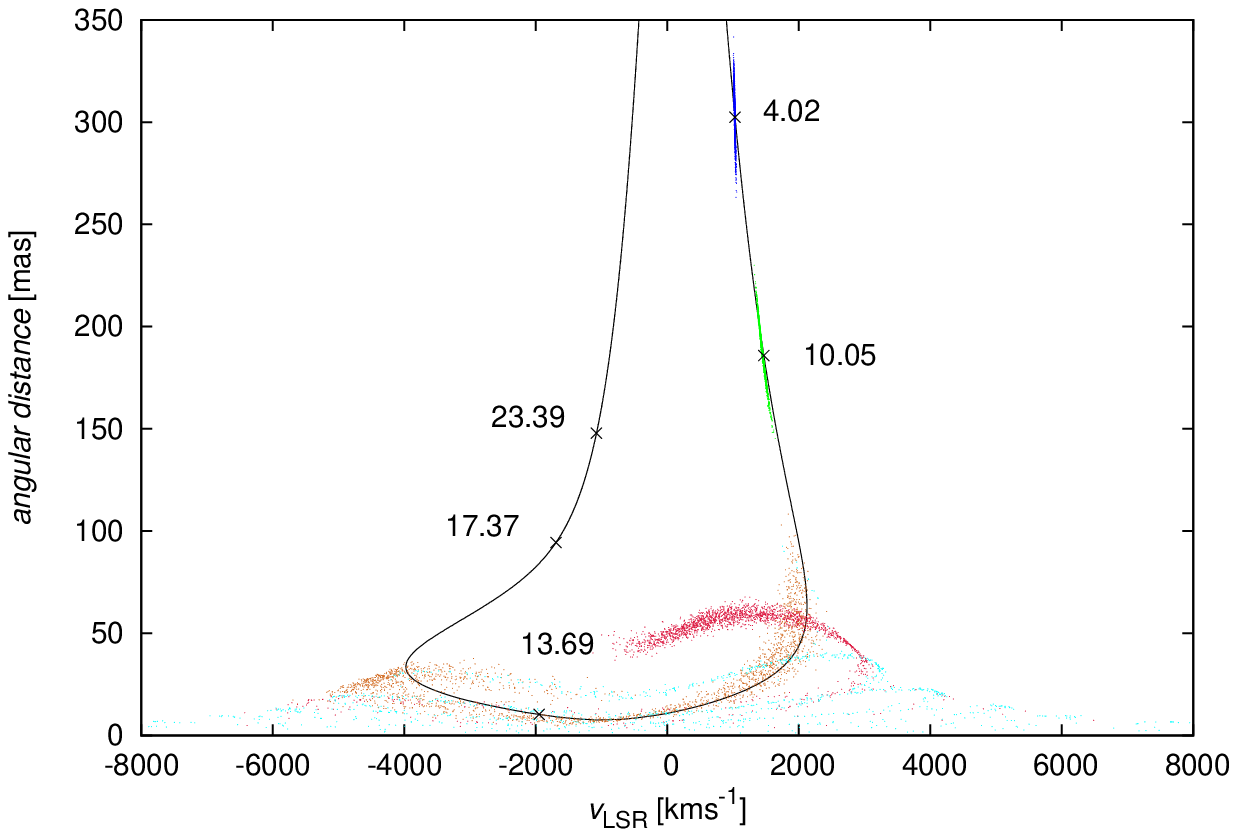}
  \includegraphics[width=0.49\textwidth]{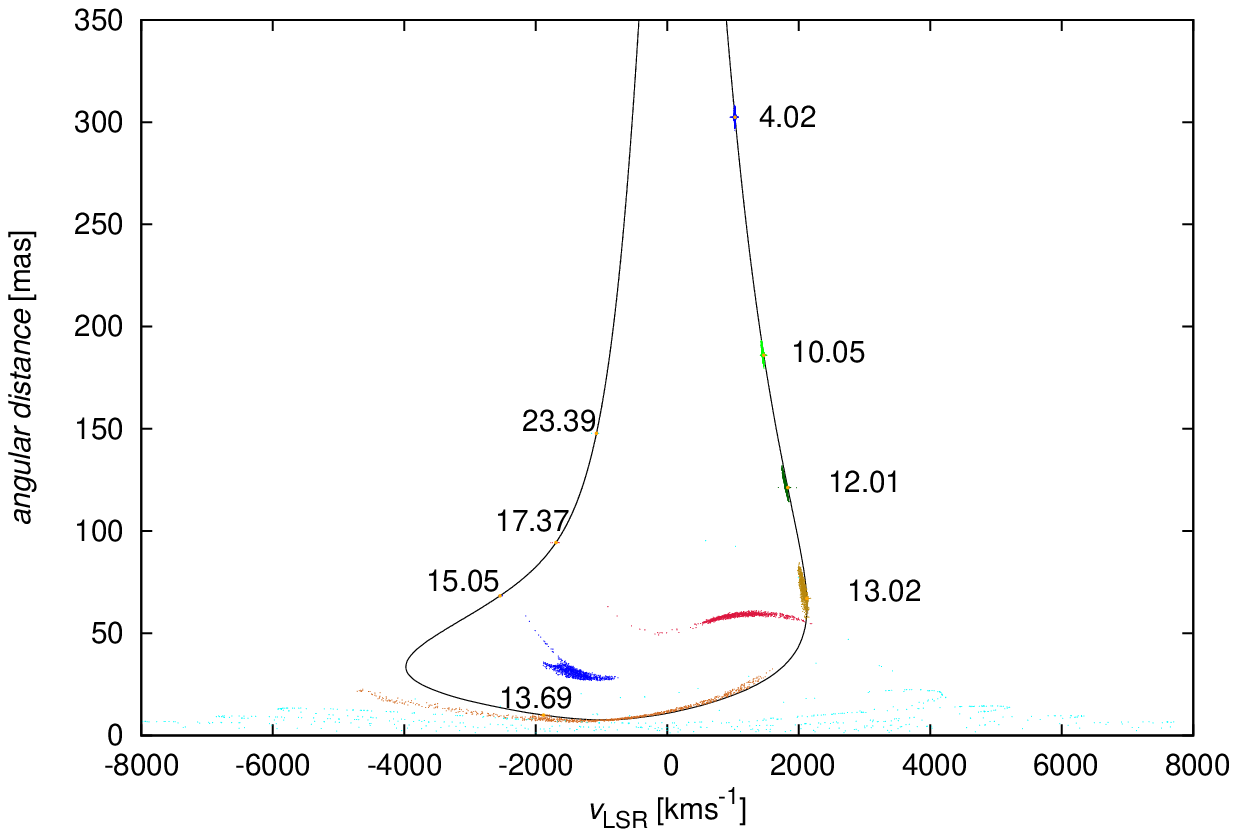} 
\caption{Numerical representation of the dust/gas cloud in terms of elementary parcels, trajectories of which are integrated
in the gravitational field and influenced by the hydrodynamical drag. 
\textit{Top left:} the evolution of the cloud in the orbital plane for selected epochs since the start of the simulation. 
Snapshots of the cloud are distinguished by different colours and shown at several moments of time (time marks 
correspond to years since the initial moment of integration). The cloud particles are characterised by the Gaussian 
dispersion around the nominal trajectory in the phase space (the black curve). \textit{Bottom left:} representation 
of the evolution from the top--left panel is shown in the velocity--distance plane. The velocity is transformed 
into the local standard of rest. The angular distance is expressed in miliarcseconds.
\textit{Top right:} the evolution as in the top-left panel, but for a cloud with an embedded star in the core. 
\textit{Bottom right:} the track in the velocity--position plane corresponding to the trajectory from the top--right panel. 
}
\label{img_cloud510-6_distvel}
\end{figure*}

The profiles in eq.\ \eqref{density_profile} were originally derived for the accretion flows up to $\sim100\,r_{\rm{s}}$, 
whereas we considered the pericentre passages at a typical distance one order of magnitude larger, $r\sim10^3\,r_{\rm{s}}$. 
However, the densities inferred from the one-dimensional model 
\citep[used to fit Chandra X-ray data farther away from the Galactic centre;][]{2004ApJ...613..322Q} 
do not differ much when extrapolated from the inner flow 
to the region of our interest \citep[factor of $\sim3$; see][for comparisons]{2012ApJ...759..130P}.    

 In general, the equation describing the motion of dust particles in the gaseous environment (optically thin stellar atmospheres, protoplanetary discs) is
\begin{equation}
\frac{\mathrm{d}\mathbf{v}}{\mathrm{d}t}=\mathbf{a_{\rm{grav}}}+\mathbf{a_{\rm{drag}}}+\mathbf{a_{\rm{rad}}}\,,
\label{eq_motion}
\end{equation} 
where $\mathbf{a_{\rm{grav}}}$ represents gravitational interactions, $\mathbf{a_{\rm{drag}}}$ stands for the wind drag acceleration, and $\mathbf{a_{\rm{rad}}}$ is the acceleration due to radiation pressure. 

The radiation pressure acceleration acting on a grain that is located at distance $r_{\star}$ from the star with flux $\Phi_{\star}=L_{\star}/(4\pi r_{\star}^2)$  may be expressed by (e.g., \citeauthor{2003ASSL..293.....B},  \citeyear{2003ASSL..293.....B})

\begin{equation}
\mathbf{a_{\rm{rad}}}=Q\frac{\Phi_{\star} \sigma_{\rm{eff}}}{mc}\left[\left(1-\frac{\mathbf{r_{\star}}\cdot\mathbf{v_{\star}}}{r_{\star}c}\right)\frac{\mathbf{r_{\star}}}{r_{\star}}-\frac{\mathbf{v_{\star}}}{c}\right]+\mathcal{O}\left(\frac{v_{\star}^2}{c^2}\right)\,,
\label{eq_rad_pressure}
\end{equation}  
where the factor $Q$ reflects the way the particle absorbs or reflects light, $m$ denotes the dust mass, $\sigma_{\rm{eff}}$ its effective cross-section, $\mathbf{v_{\star}}$ is the velocity vector in the frame of the star, and $c$ is the speed of light. The terms in the brackets represent the direct pressure, the change of radiation energy due to the Doppler effect, and the Poynting-Robertson drag, respectively. Eq. \eqref{eq_rad_pressure} is valid in a Newtonian approximation that is consistent with the pericentre passage of a star at the distance of $\sim 1000\, r_{\rm{s}}$ at most. However, in the immediate vicinity of the SMBH, it would be necessary to include the second and higher powers of $(v_{\star}/c) $ as well.

 The first two radial terms in eq. \ref{eq_rad_pressure} effectively
change the mass of the star, which itself is a free parameter in our model, and the non-radial Poynting-Robertson term causes the decrease in the semimajor axis and eccentricity of the orbit and contributes to the inspiral of dust towards the star on the Poynting-Robertson time scale $\tau_{\rm{pr}}$, which may be approximated using eq. \eqref{eq_rad_pressure} and setting the effective cross-section equal to the geometrical cross-section of the grain,  $\sigma_{\rm{eff}}=\pi R^2$:

\begin{align}
\tau_{\rm{pr}} &= \frac{16 \pi c^2}{3 Q}L_{\star}^{-1}R \rho_{\rm{d}} r^2 \notag \\
               &= 28 \times 10^2\,\rm{yr}\, \frac{1}{Q}    \left(\frac{L_{\star}}{L_{\odot}}\right)^{-1}\left(\frac{R}{1\,\rm{\mu m}}\right) \left(\frac{\rho_{\rm{d}}}{1\,\rm{gcm^{-3}}}\right)\left(\frac{r}{1\,\rm{AU}}\right)^2. 
\label{eq_pr_timescale}               
\end{align}
In our model, we worked with  time scales of $100\,\rm{yr}$, which are shorter than $\tau_{\rm{pr}}$ by one order of magnitude for distances of $\sim 1\,\rm{AU}$. For more distant orbits, the time scale is significantly prolonged because of its quadratic dependence on the distance.

The radiation pressure from other S-stars in the background also contributes, but it is at least two orders of magnitude weaker.  For an estimate, we took $\sim 500\,\rm{mas}\approx 4125\,\rm{AU}$ as an average distance of individual S-cluster members. The estimated number of $\sim 20$-$M_{\odot}$ stars is $\sim 30$ and we compared the effect of their radiation with a low-mass (LM) star of $2\,\rm{M_{\odot}}$. Hence, the ratio of accelerations due to radiation, eq. \eqref{eq_rad_pressure}, is

\begin{align}
\frac{a_{\rm{LM}}}{a_{S}} & \approx \frac{L_{\rm{LM}}}{30L_{\rm{S}}}\left(\frac{r_{\rm{S}}}{r_{\rm{LM}}}\right)^2 \notag\\
                          & \approx \frac{1}{30}\left(\frac{M_{\rm{LM}}}{M_{\rm{S}}}\right)^{3.5}\left(\frac{r_{\rm{S}}}{r_{\rm{LM}}}\right)^2 \notag\\
                          & = \frac{1}{30} \left(\frac{2\,\rm{M_{\odot}}}{20\,\rm{M_{\odot}}}\right)^{3.5}\left(\frac{4125\,\rm{AU}}{1\,\rm{AU}}\right)^2 \approx 180\,. 
\end{align}
Similarly, the Poynting-Robertson time scale, eq. \eqref{eq_pr_timescale}, yields $\tau_{\rm{pr}} \approx 10^5\,\rm{yr}$ for the average distance of $500\,\rm{mas}$ and the luminosity of $\sim 30$ $20$-$M_{\odot}$ stars. \citet{2013ApJ...768..108S} also estimated that although a production rate of Lyman continuum photons is high, very few of them are intercepted because of the small area of the bow shock formed ahead of the star. Hence, the effect of radiation pressure from background stars is weak on the time scale of $\sim 100\,\rm{yr}$ considered here.

 The radiation drag in the vicinity of a putative star will only
slightly affect
 individual orbits of grains on the time scale of a few orbital periods. The ensemble of particles behaves similarly with and without the radiation term involved, which shows that gravitational forces and hydrodynamical drag are dominant for their dynamics. To reduce computation costs, we omitted this term in most of our numerical calculations.

The acceleration caused by the drag for both supersonic (Stokes law) and subsonic (Epstein law) modes may be written in a closed form (e.g.,                    
 \citeauthor{1975ApJ...198..583K}, \citeyear{1975ApJ...198..583K}, \citeauthor{2011ApJ...734L..26V}, \citeyear{2011ApJ...734L..26V}):

\begin{equation}
\mathbf{a_{\rm{drag}}}=-\eta_{\rm{drag}} \frac{\pi R^2}{m} \rho_{\rm{a}} \sqrt{v_{\rm{rel}}^2+\overline{v}_{\rm{t}}^2}\,\mathbf{v_{\rm{rel}}}\,,
\label{eq_drag_term} 
\end{equation}

where $\mathbf{v_{\rm{rel}}}=\mathbf{v}-\mathbf{v_{\rm{g}}}$ is the relative velocity of grains with respect to the gas motion $\mathbf{v_{\rm{g}}}$. The mean thermal velocity, assuming Maxwellian distribution, is $\overline{v}_{\rm{t}}=(8/\pi)^{1/2}c_{\rm{s}}$, which is proportional to the sound speed $c_{\rm{s}}=\sqrt{kT/m_{\rm{H}}}$ (for ideal gas). The factor $\eta_{\rm{drag}}$ acquires the value $4/3$ for the Epstein law and $1/2 C_{\rm{d}}(Re)$ for the Stokes law, where the coefficient $C_{\rm{d}}(Re)$ depends on the Reynolds number $Re$ (for an in-depth discussion see \citeauthor{2011ApJ...733...56P}, \citeyear{2011ApJ...733...56P}). As shown in Fig. \ref{img_velocity}, orbital velocities within the inner $\sim 10^6$ Schwarzschild radii are supersonic,  $v_{\rm{orb}} \approx \sqrt{GM_{\bullet}/r} > c_{\rm{s}}$, so the Stokes law is approximately valid. This is also true for grains embedded in the outflowing wind atmosphere close to the star, see Fig. \ref{img_orbvel_cs}. If we assume an isothermal atmosphere with $T\approx 10\,000\rm{K}$ that is in ionisation equilibrium with the environment, orbital velocities are approximately equal to the sound speed at distance $r_{\rm{s}}\approx G M_{\star} m_{\rm{H}}/(k T)$ and decrease. However, relative velocities with respect to radial outflow, in case of circular orbits of grains, are higher than the sound speed by about factor of 10 for typical terminal wind velocities of $100\,\rm{km\,s^{-1}}$. Assuming that grains have a spherical shape with diameter $d$ and substituting the dust grain volume times density for the mass, we obtain $\mathbf{a} \approx -\frac{3}{4} C_{\rm{d}}(Re)(\rho_{\rm{a}}/\rho_{\rm{d}})d^{-1}v_{\rm{rel}}\mathbf{v_{\rm{rel}}}$ and we set $C_{\rm{d}}=1$.\footnote{From subsonic to supersonic regimes, the drag coefficient changes typically by about a factor of two.}
 
 We note that the magnetic field frozen in the stellar wind and potentially that of the interstellar medium near the galactic nucleus as well may influence the motion of charged dust, mainly of a smaller size. This is mainly determined by the ratio of the charge of the grain $q$ resulting from photo-charging and its mass $m$, $q/m$. When the main component of the field is radial, $\mathbf{B}=B_{\rm{r}}\hv{r}$, and dust motion is azimuthal, $\mathbf{v}=v_{\phi}\hv{\boldsymbol\phi}$, the Lorentz acceleration, $a_{\rm{L}}=(q/m) v_{\rm{\phi}} B_{\rm{r}}$ causes the grain inclination to increase or decrease, depending on the mutual orientations of the magnetic field and velocity vectors. The field orientation may flip after some time and may lead to oscillations around the orbital plane. Because of the many uncertainties concerning the magnetic field strength and the efficiency of photo-charging, we did not consider the magnetic field in our analysis.
 
   Despite the arguments for neglecting these terms and effects,  this is just a crude model. Nonetheless, it allows us to reproduce
the relevant trends in a semi-analytical way.

In figure \ref{img_comparison_acceleration} we plot the ratio of the magnitude of the acceleration 
due to the drag force \eqref{eq_drag_term} with respect to the gravitational acceleration 
across the range of radii. In this case we neglected the motion of ambient 
medium and set the relative velocity $v_{\rm{rel}}$ equal to the orbital speed.
Although the ratio varies significantly (by six orders of magnitude), it is non-negligible for 
some typical particle sizes in the range $\sim$\,$0.1\,\mu\rm{m}$--$1\,\mu\rm{m}$
(for the smallest particles it exceeds the gravitational acceleration magnitude).\footnote{The 
effect might be more significant in AGN, where the typical density of the environment is 
higher than in inactive nuclei, but in that case strong irradiation leads
to dust heating and destruction in a relatively large volume around the central source.} 

As the cloud approaches the central region, the dust absorbs the radiation mainly at shorter wavelengths 
and heats up. The dust temperature grows to the sublimation temperature $T_{\rm{sub}}$ and eventually leads to sublimation. 
When we take into account only the radiation of the central accretion flow, the sublimation radius 
is easily determined \citep[e.g.,][]{1987ApJ...320..537B}:
\begin{equation}
r_{\rm{sub}} = 567\, T_{\rm{sub}}^{-2.8}\left[\left(\frac{L_{\rm{uv}}}{L_{\odot}}\right)\exp{(-\tau_{\rm{uv}})}\right]^{\frac{1}{2}}\, \rm{[pc]},
\label{eq_sub_radius}
\end{equation}    
where $L_{\rm{uv}}$ stands for the luminosity of the central source at ultraviolet wavelengths 
expressed in units of solar luminosity $L_{\odot}$. The optical depth, $\tau_{uv}$, enables us to 
estimate the sublimation radius or inversely the temperature of dust at given $r$
\citep[again, realistic profiles must be more complicated;][]{2011A&A...536A..78K}. Furthermore, in 
sources with radiatively efficient accretion, X-ray heating contributes significantly \citep{2011A&A...525L...8C}.

For the sublimation temperature $T_{\rm{sub}} \approx 1\,500\,\rm{K}$ \citep{1987ApJ...320..537B}, 
low optical depth, and $L_{\rm{uv}} \approx 10^3\,L_{\odot}$, the relation 
\eqref{eq_sub_radius} gives $r_{\rm{sub}} \approx 2.3 \times 10^{-5} \rm{pc} \approx 56\,r_{\rm{s}}$ as a
lower estimate (we did not consider hot OB stars present near the centre, whose radiation probably heats the dust further). When the stellar populations are taken into account, the central luminosity 
can be estimated to a few $\sim\,10^6\,L_{\odot}$ for the smooth distribution, and up to 
$10^7\,\,L_{\odot}$ for the clumpy structure \citep{1992ApJ...387..189D}, which leads to 
$r_{\rm{sub}}\approx 2.3 \times 10^{-3} \rm{pc}\approx 5600\,r_{\rm{s}}$ as an upper limit for the sublimation 
radius. 

Given the uncertainty, it is appropriate to refer to the sublimation zone extending from 
a few tens of Schwarzschild radii up to $\sim 5\times10^3\,r_{\rm{s}}$. To shorten the 
integration time of numerical experiments, we safely set the inner sublimation radius 
to $300\,r_{\rm{s}}$ in our test runs. Using the inverse relation $T=T(r,\tau_{uv})$ in 
eq. \eqref{eq_sub_radius}, we estimate the temperature profile of dust near the source 
of central luminosity $10^5\,L_{\odot}$, which is intermediate between the limits of 
$10^3$ and $10^7\,L_{\odot}$ (see figure~\ref{img_tempcomp}).

The passages with pericentre above $\simeq1\times10^3\,r_{\rm{s}}$ (the case of G2/DSO, 
the Dusty S-cluster Object in the Galactic centre)\footnote{The S-cluster is the $1^{\prime\prime}$ 
diameter star cluster of high-velocity stars surrounding Sgr A* 
\citep{1996Natur.383..415E,1998ApJ...509..678G,2005bhcm.book.....E,2007gsbh.book.....M,2010RvMP...82.3121G}.}
probably does not heat the dust to the sublimation temperature, which would only be reached 
at $\approx 600\,r_{\rm{s}}$. However, the mechanism of heating is more complicated not only 
because of the presence of stars, but also because of the shock heating that may contribute 
in case of transonic motion.
  
\subsection{Role of central star}
 It has often been advocated (e.g., \citeauthor{2013ApJ...774...44G}, \citeyear{2013ApJ...774...44G}) that the infalling DSO/G2 source may represent a core-less clump that has originated farther out at greater distance, $\gtrsim10^4r_{\rm{}s}$, 
and now is on the way toward 
the pericentre, where it will largely disintegrate and undergo accretion onto the SMBH. However, it may be also a dusty
envelope that enshrouds a star. Extended shells surrounding stars near Sgr~A* have been reported, see for example \citet{2005A&A...443..163M}.

Before the pericentre passage, the two cases are expected to produce a similar image and the
corresponding spectrum to contain a thermal component from the dust. On the other hand, a transit through the
pericentre must reveal the nature in a clearer way. Because we considered a pericentre distance of the order of 
$10^3r_{\rm{}s}$, the star itself is not destroyed by tidal forces of the SMBH, instead, it continues to follow the eccentric
ellipse. The diluted atmosphere is more visibly affected by its interaction with the ambient medium.

First, we compared the temporal evolution of the spatial distance and velocity with/without the hydrodynamical drag 
for the high-eccentricity orbits of a single particle. As a specific example, we used the L'-band based nominal orbital elements of  
G2/DSO  \citep{2013ApJ...763...78G}. The resulting evolution is plotted in figure~\ref{img_dbh_v3d}; it
appears to be consistent with eccentricities and semi-major axes of dust grains gradually decreasing, as they 
move in the ambient medium. 

Next, we considered a cloud of particles with a certain initial distribution of positions and velocities and  included 
different effects acting on the motion. \citet{2012Natur.481...51G,2013ApJ...763...78G} studied the evolution of such a cloud 
of particles in the field of the Sgr~A* SMBH. We revisited their simulations by including the interaction with different prescriptions
for the ambient wind. 
In our simulations, for example, 2000 particles were initially distributed according to a spherically symmetric Gaussian 
distribution in the position--velocity phase space. We used the same L'-band-based orbital elements for 
the G2/DSO cloud as above. 

The evolution in the orbital plane and in the velocity--distance plane are plotted in figure~\ref{img_cloud510-6_distvel} 
(left panels) for selected epochs since the start of the simulation. The initial FWHM of the phase-space distribution was 
set to be $25\,\rm{mas}$ for positions \citep[the value adopted from][]{2013ApJ...774...44G}, and $5\,\rm{km/s}$ as a 
typical turbulent velocity within the cloud. 
These simulations reveal that because of the drag of the ambient ISM and gravity of the SMBH, most particles 
continue to spiral in towards the centre. The cloud as a whole becomes progressively stretched by the tidal forces, 
and the velocity dispersion rises abruptly as particles approach the SMBH.

Secondly, we examined the evolution with an embedded star, whose mass has been set to $M_\star=2M_{\odot}$. Dust grains ($d=5\,\rm{\mu m}$)
initially obey a uniform distribution of semi-major axes in the range $(0.1,50)\,\rm{AU}$, eccentricities in the range 
$(0,0.1)$, and inclinations within $(0^{\circ},180^{\circ})$.\footnote{Inclination values lower than $90^{\circ}$ correspond
to a prograde (direct) orbit, values exceeding $90^{\circ}$ are reserved for the retrograde sense of motion.} The distribution of velocities is Keplerian. Fig.\ 
\ref{img_cloud510-6_distvel} also shows the evolution of such a stellar source (right panels).

The particles start to trail behind the star because of the drag. In combination with the tidal effects, the position and 
the velocity dispersion gradually increase up to the pericentre. During the post-pericentre phase the 
velocity dispersion first decreases, then increases again while particles spiral inwards to 
the SMBH. In Fig. \ref{img_cloud510-6_distvel} we clearly see how the attraction of the star influences 
some of the particles (cp.\ the epoch $15.05$). A small fraction of particles remain bound to the star; 
specifically for this run, about $5\permil$ particles remained bound to the star; this number did not 
change after the second pericentre passage. 

\subsection{Dust truncation radius}
\label{truncation_radius}

To estimate the dust mass diverted from the initial nominal trajectory of the cloud at its pericentre passages, it is useful to consider the 
Hill radius of the binary system star--SMBH. At the pericentre,

\begin{equation}
r_{\rm{H}}= a(1-e)\left(\frac{M_{\star}}{3 M_{\bullet}}\right)^{1/3}\,,
\label{eq_hill_rad}
\end{equation}  
which for the mass $M_{*}=2{M_{\odot}}$ and the position at the pericentre yields $\sim 1\,\rm{AU}$; inside 
this sphere of influence the particles remain bound to the star, while outside it the particles are captured by the SMBH. 

\begin{figure}[tbh]
\centering
 \includegraphics[width=0.5\textwidth]{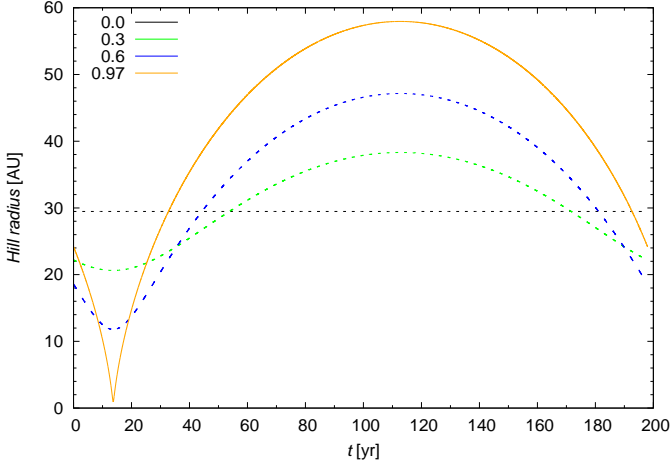}
\caption{Critical Hill radius $r_{\rm{H}}(t)$ for four different eccentricities: $e=0.0$, $0.3$, $0.6$, and $0.97$.
Enhanced accretion occurs when some of the dust particles move
beyond $r_{\rm{H}}$.}
\label{img_hill_rad}
\end{figure}

The sphere of influence increases after each pericentre passage, as illustrated in figure~\ref{img_hill_rad}, 
where we plot the Hill radius evolution over an interval of time. We varied the eccentricity, but other orbital 
parameters remained fixed ($M_\bullet=4.4\times10^6M_\odot$ for Sgr~A* SMBH). Highly eccentric trajectories 
($e \gtrsim 0.9$) are prone to a significant mass loss when the atmosphere of the star undergoes the Roche-lobe overflow onto the black hole. 

\begin{figure}[h!]
\centering
\includegraphics[width=0.5\textwidth]{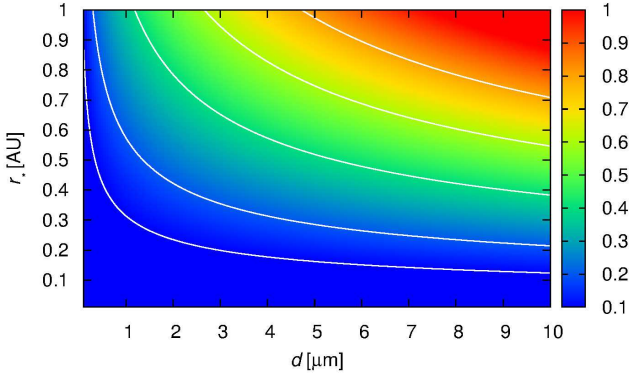}
\caption{Colour-coded plot of the wind-truncation radius (colour bar units in AU). Each point in the graph estimates the truncation radius from the equilibrium between gravitational and drag forces for a given distance from the star and a grain diameter. The contours stand for values $0.1$, $0.2$, $0.4$, $0.6$, and $0.8\,\rm{AU}$.}
\label{img_wind_truncation}
\end{figure}

\begin{figure}[tbh]
\centering
 \includegraphics[width=0.5\textwidth]{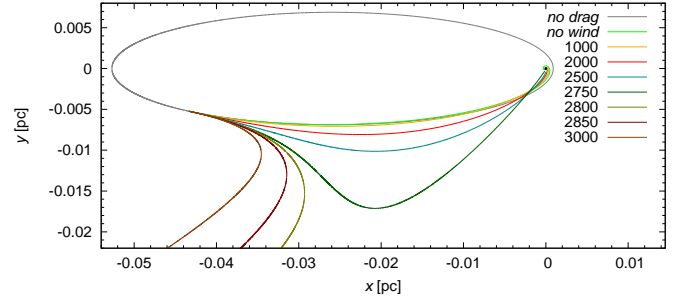}
\caption{Trajectory of a single dust grain ($d=1\,\rm{\mu m}$) for different cases of the central wind outflow velocity. The grey ellipse corresponds
to the nominal trajectory, neglecting the drag by the wind. The values of the outflow velocity are expressed in $\rm{km/s}$
in the legend.}
\label{img_1tp_wind}
\end{figure}   

 The Hill radius expressed by eq. \eqref{eq_hill_rad} employs only tidal shearing. However, in our case the motion of dust is influenced by drag from the wind outflow, and this may indeed decrease the critical radius at which orbits of grains become unstable and consequently leave the circumstellar environment. Following the analysis of \citet{2011ApJ...733...56P}, we define the wind-truncation radius at which the gravitational acceleration acting on the grain $GM_{\star}/r_{\star}^2$ is equal to the acceleration $a_{\rm{drag}}$ resulting from the wind drag:

\begin{equation}
r_{\rm{w}}=\left(\frac{GM_{\star}}{a_{\rm{drag}}}\right)^{1/2}\,.
\label{eq_wind_radius}
\end{equation}
After inserting the drag term \eqref{eq_drag_term} into \eqref{eq_wind_radius}, we obtain

\begin{equation}
r_{\rm{w}}=\left(\frac{GM_{\star}m}{\eta_{\rm{drag}}\pi R^2 \rho_{\rm{\star}} (v_{\rm{rel}}^2 +\overline{v}_{\rm{t}}^2)^{1/2} v_{\rm{rel}}}\right)^{1/2}\,.
\label{eq_wind_radius2}
\end{equation}

To derive an estimate, we assumed the Stokes law and rewrote eq. \eqref{eq_wind_radius2} as a function of grain diameter $d$ and distance from the star $r_{\star}$,

\begin{equation}
r_{\rm{w}}(d,r_{\star})=\left(\frac{4}{3}\frac{GM_{\star}}{C_{\rm{d}}}\rho_{\rm{d}}\right)^{1/2}\left(\frac{d}{\rho_{\star}v_{\rm{rel}}^2}\right)^{1/2}\,,
\label{eq_wind_radius3}
\end{equation} 
 where both the stellar wind density $\rho_{\star}$ and the relative velocity of grains with respect to the wind $v_{\rm{rel}}$ are functions of the distance from the star. We rewrote eq. \eqref{eq_wind_radius3} into a convenient form, taking $\rho_{\rm{d}}=2\,260\,\rm{kg\,m^{-3}}$ for the density of dust (e.g., \citeauthor{1987ApJ...320..537B}, \citeyear{1987ApJ...320..537B}) and assuming the density of spherical wind outflow $\rho_{\star}=\dot{m}_{\rm{w}}/(4\pi r_{\star}^2 v_{\rm{w}})$,
 
\begin{align}
r_{\rm{w}}=0.28\,\left(\frac{M_{\star}}{M_{\odot}}\right)^{1/2}\left(\frac{d}{\mu m}\right)^{1/2}\left(\frac{v_{\rm{rel}}}{100\,\rm{km\,s^{-1}}}\right)^{-1}
\left(\frac{v_{\rm{w}}}{100\,\rm{km\,s^{-1}}}\right)^{1/2} \notag\\
\times \left(\frac{\dot{m}_{\star}}{10^{-8} M_{\odot} yr^{-1}}\right)^{-1/2}\left(\frac{r_{\star}}{\rm{AU}}\right)\,\rm{AU}\,.
\label{eq_wind_radius4}
\end{align}

In Fig. \ref{img_wind_truncation}, we plot the wind-truncation radius as a function of both the grain diameter (in $\mu m$) and the distance from $2\,\rm{M_{\odot}}$-star (in $AU$) with the typical mass-loss rate of $10^{-8}\,\rm{M_{\odot}\,yr^{-1}}$ and the terminal wind speed of $100\,\rm{km\,s^{-1}}$. The wind drag generally causes the stability region to shrink below the Hill radius (in our case $\sim 1\,\rm{AU}$ at the pericentre), mainly affecting small grains $\lesssim 1\,\rm{\mu m}$.

\section{Results of simulations}
\label{sec2}
\subsection{Effects of wind and a bow shock -- the case of G2/DSO}
Comet-shaped features have been observed near the Galactic centre that point to the role of the radial outflow 
in the form of a wind from the central region \citep{2010A&A...521A..13M}. The source of the wind can be hot 
stars and/or the accretion flow in the vicinity of the SMBH.  A similar effect  is expected to occur in our model as well. One of the differences that distinguishes the core-less cloud from a dust-enshrouded star is 
the size of the bow shock \citep[e.g.,][]{2011ApJ...734L..26V,2013MNRAS.436.3626A}. In the case of a star, the bow shock is formed by the wind interaction; a
powerful wind outflow (as in the case of massive stars) develops the stagnation point radius at a much larger 
radius than the size of the star. On the other hand, in the case of a core-less cloud the bow shock is only slightly 
larger than the size of the cloud.

\begin{figure*}[tbh]
\centering
 \includegraphics[width=0.52\textwidth]{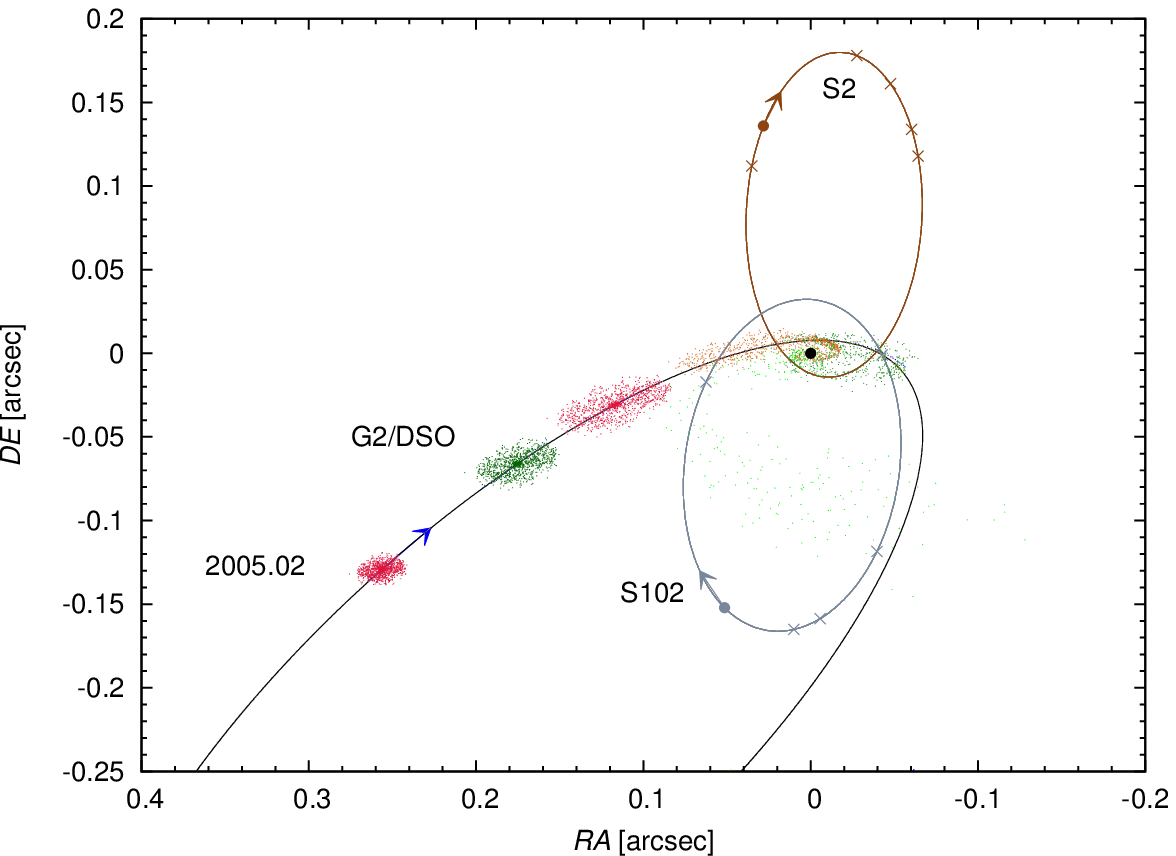}
 \hspace*{5mm}
 \includegraphics[width=0.40\textwidth]{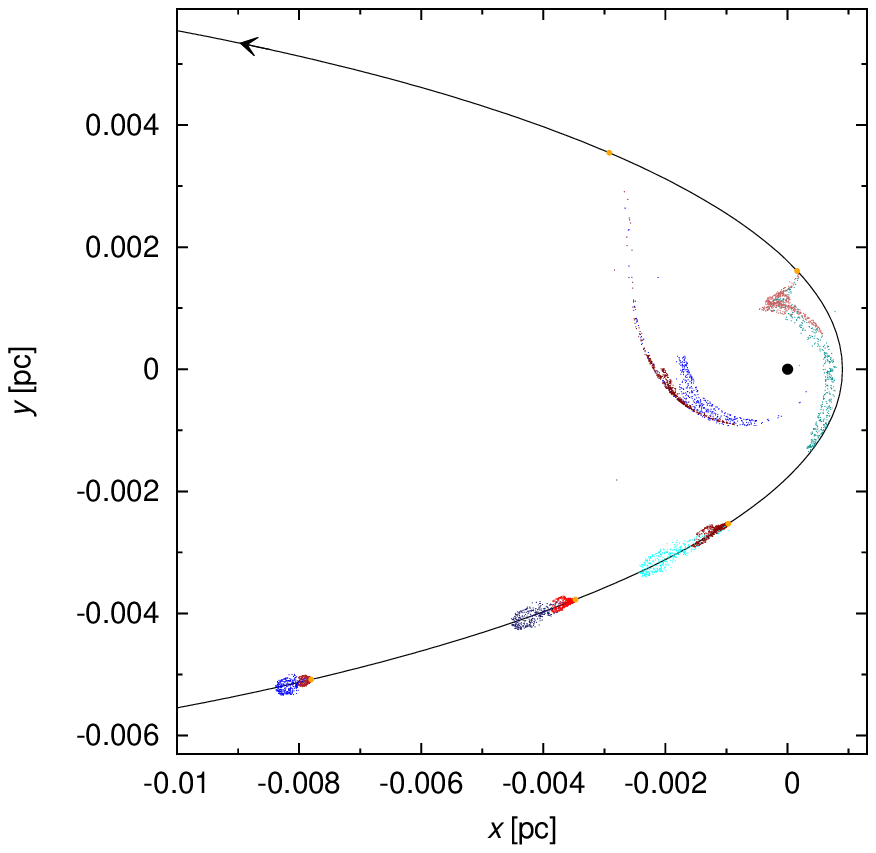}
 \caption{\textit{Left panel:} a sketch of the orbit orientation on the sky, consistent with 
 the current nominal trajectory of the G2/DSO (i.e., arriving from the south-east direction and going into pericentre from the 
 north, then leaving towards the south-east, showing the offset from the SMBH position in the origin). Two eccentric orbits of 
 the S-stars are also indicated. \textit{Right panel:} two cases of the possible evolution a gaseous/dusty stellar envelope, 
 initially bound to the star: (i) with a radial wind blowing from the centre ($v_{\rm w}=1000\;\mbox{km/s}$; bluish colours), 
 and (ii) without the outflowing wind (reddish colours). Here the same nominal trajectory
 as in the left panel is plotted, now with respect to the $(x,y)$ orbital plane. See the text for details.}
\label{img_stardust_wind/nowind}
\end{figure*} 

We included the radial wind outflow by setting 
$\mathbf{v}_{\rm{rel}}=\mathbf{v}-\mathbf{v_{\rm{w}}}$ in eq.\ \eqref{eq_drag_term}, where $\mathbf{v}$ stands for the velocity 
of numerical particles with respect to the SMBH, and $\mathbf{v_{\rm{w}}}$ is the wind velocity (we assumed a 
spherical outflow at $r\lesssim 1''$). The particles feel the wind drag; the computed trajectories for particles of 
$d=1\,\rm{\mu m}$ are shown in figure~\ref{img_1tp_wind}. The initial conditions were the same for all cases and were computed according to the nominal orbital elements of the G2/DSO 
object. For small and moderate wind velocities ($\lesssim 1000\,\rm{km/s}$), the trajectory is only slightly affected. The time of pericentre passage is delayed and the rate of spiralling inwards increases
because there is an additional non-radial term (due to the aberration). For $v_{\rm{w}}\gtrsim 2800\,\rm{km/s}$
the particles are blown away with the wind. \citet{2010A&A...521A..13M} found evidence of the presence of wind
speed from the centre that are of this order of magnitude.

Consequently, we compared the evolution of a bound star--gas--dust system with and without the effect of a spherical 
wind from the centre. Particles ($d=2.5\,\rm{\mu m}$) were initially distributed uniformly around the star 
$(M_\star=3{M_{\odot}}$ in this example). The range of inclination was set from $0^{\circ}$ to $180^{\circ}$, semi-major axes spanned
the interval $(0.001,10)\,\rm{AU}$, and eccentricities varied from $0$ to $0.1$. The resulting evolution is 
plotted in figure~\ref{img_stardust_wind/nowind}. The wind slows the particles down and the initial shape of the shell is more 
stretched than in the no-wind case.

If a wind-blowing star moves supersonically through ISM, a bow shock is formed.
The effect has also been confirmed at $0.2\,\rm{pc}$ distance
from the Galactic centre \citep{2010A&A...521A..13M}. Here we employed a two-dimensional model
of the bow shock,
\begin{equation}
R(\theta)=R_0 \csc{\theta}\sqrt{3(1-\theta\cot{\theta})}\,,
\label{eq_bowshock_shape}
\end{equation} 
which we rotated around the symmetry axis to obtain the correct orientation of the shock in 3D. In eq.\
\eqref{eq_bowshock_shape}, $R_0$ stands for the stand-off distance, where the ram pressures of the 
ambient medium and the stellar wind are at balance \citep{1996ApJ...459L..31W},
\begin{equation}
R_0=\left(\frac{\dot{m}_{\rm{w}} v_{\rm{w}}^{\star}}{4\pi \rho_{\rm{a}} v_{\rm{rel}}^2}\right)^{1/2}\,,
\label{eq_standoff_distance}
\end{equation}  
where $\dot{m}_{\rm{w}}$ is the stellar mass-loss rate, $v_{\rm{w}}^{\star}$ is the terminal velocity of the stellar wind. 
We used eq.\ \eqref{density_profile} to find $\rho_{\rm{a}}$, and $v_{\rm{rel}}=\lvert \mathbf{v}-\mathbf{v_{\rm{w}}}\rvert$ at a specific point 
on the orbit. The symmetry axis of the bow shock is aligned with the direction of the relative velocity, 
$\mathbf{v_{\rm{rel}}}$, as expected.         

 We assumed that the wind-blowing star has a spherical wind with the constant magnitude equal to its terminal velocity. 
In our numerical scheme, particles outside the bow shock feel the drag and wind from the ambient medium near the nucleus. 
Inside the bow shock, the drag is approximately proportional to the second power of the relative velocity with respect to the 
circumstellar environment with the radial outflow, and to the circumstellar density $\rho_{\star}$, $\rho_{\star}=\dot{m}_{\rm{w}}/(4\pi r_{\star}^2 v_{\rm{w}}^{\star})$. In further examples we used the parameters relevant for a young, low-mass (T Tauri) star \citep{2013ApJ...768..108S}: $\dot{m}_{\rm{w}}=10^{-8}\,M_{\odot}\,{\rm yr^{-1}}$, $v_{\rm{w}}^{\star}=200\,\rm{km/s}$, but the procedure may be also applied to more evolved stellar types. 


\begin{figure*}[tbh]
\centering
 \includegraphics[width=0.49\textwidth]{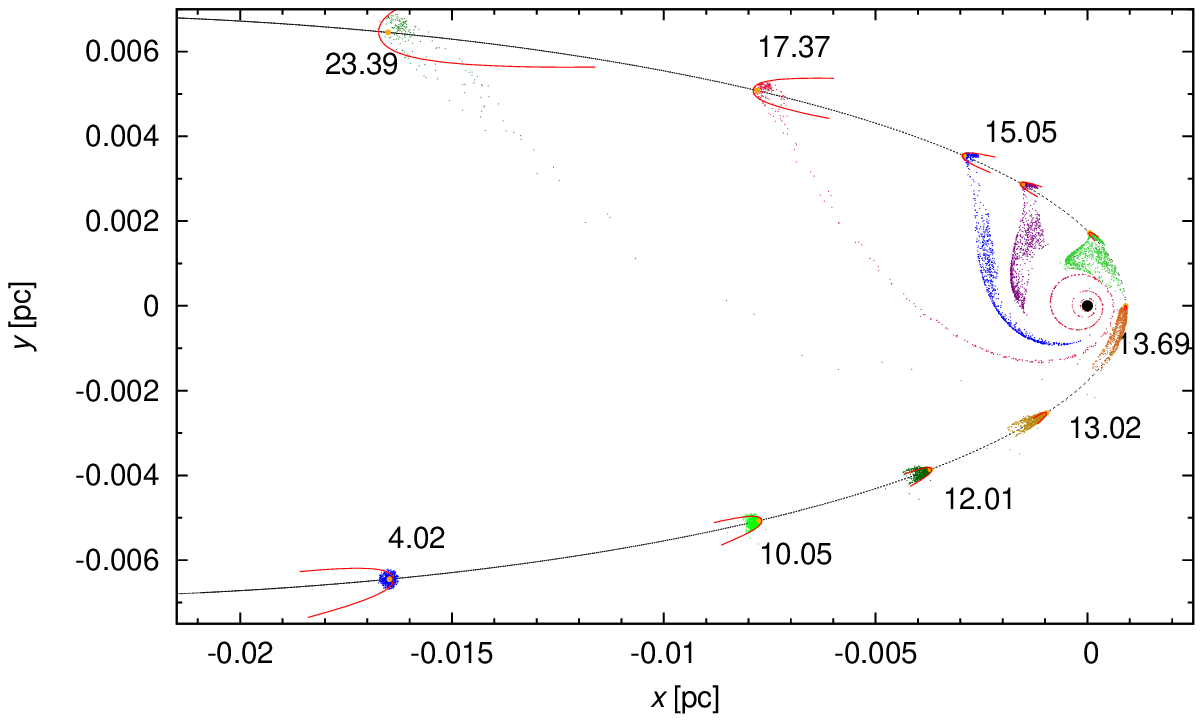}
 \includegraphics[width=0.49\textwidth]{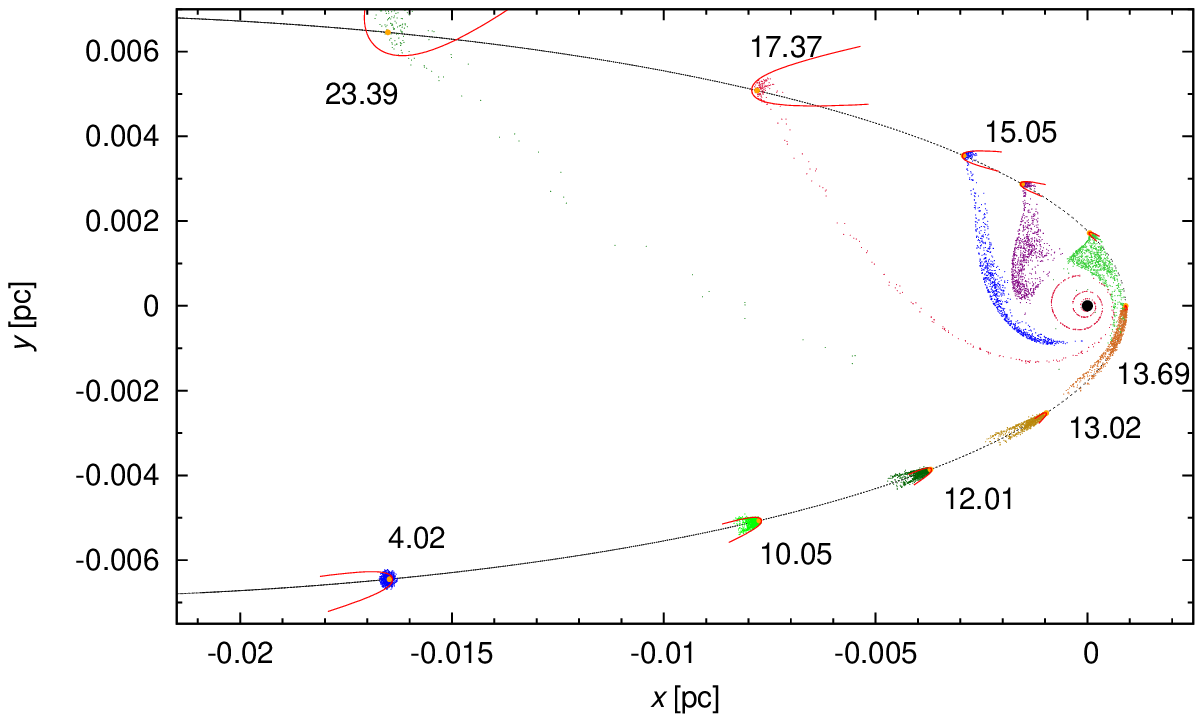}
 \includegraphics[width=0.49\textwidth]{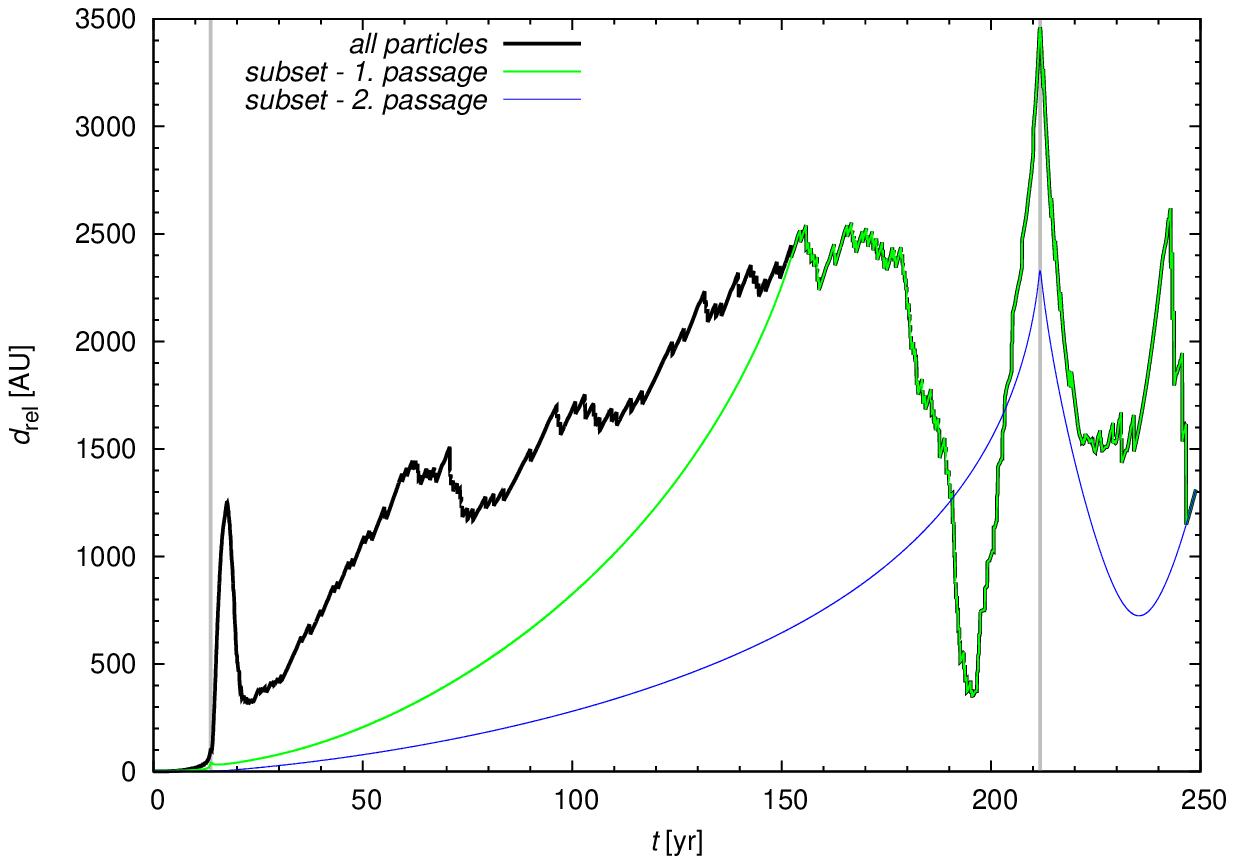}
 \includegraphics[width=0.49\textwidth]{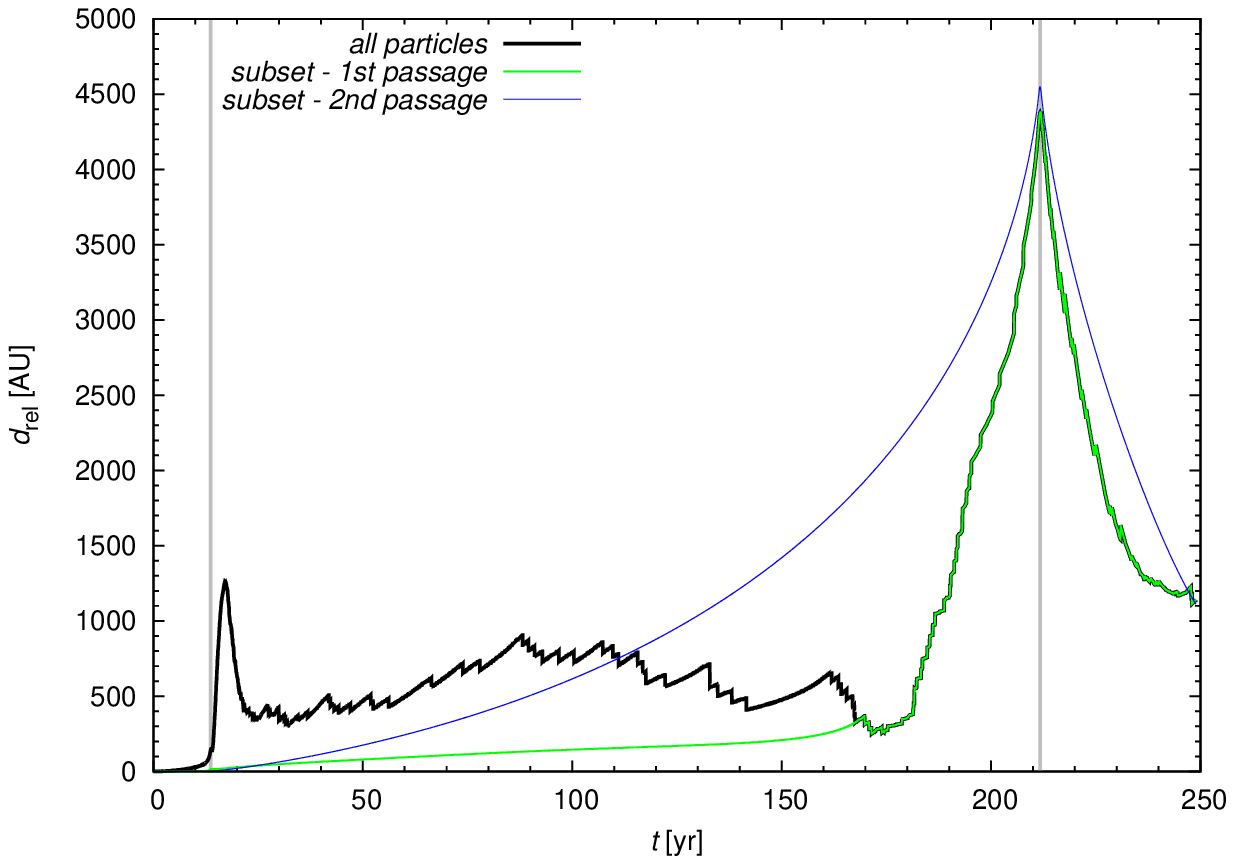}
\caption{ Evolution of the star with a dusty envelope together with an instantaneous orientation of the bow shock (top panels).
\textit{Top left:} star mass $M_\star=2{M_{\odot}}$. The velocity of the wind from the centre is 
$v_{\rm{w}}=500\,{\rm{km/s}}$. \textit{Top right:} same parameters as in the left panel except for 
$v_{\rm{w}}=1000\,\rm{km/s}$, showing the change of orientation of the bow shock. 
In both cases, the mass transfer peaks with a slight delay after the pericentre passage.
Corresponding to the two panels in the top row, an offset is shown between the star position 
along its nominal orbit from the centre of mass of the cloud in the bottom panels.
Different curves show the offset computed for all particles forming the cloud (black curve),
a subset of particles in the cloud that survive the first passage through the pericentre (green curve),
and a subset of those that also survive the second passage. Moments of pericentre
passages are indicated by vertical lines. The separation of 1000 AU corresponds to $\simeq0.12$ arcsec
at the distance to Sgr~A*.
\textit{Bottom left:}  $v_{\rm{w}}=500\,{\rm{km/s}}$. 
\textit{Bottom right:}  $v_{\rm{w}}=1000\,{\rm{km/s}}$.}
\label{img_bowshock_star_dust}
\end{figure*}          

Now we also included the effect of the bow shock on the cloud evolution. In our example we set the mass of an embedded star 
to $M_\star=2{M_{\odot}}$ and compared two cases (figure~\ref{img_bowshock_star_dust}): first, the wind blowing from the centre at velocity of $500\,\rm{km/s}$, 
and, second, the other case with $1000\,\rm{km/s}$ wind. In each, case particles have a uniform distribution of semi-major axes 
in the interval $(0.1,50)\,\rm{AU}$, inclination $(0^{\circ},180^{\circ})$, and eccentricity $(0,0.1)$. We note that the
character and orientation of the bow-shock sources can in principle be tested observationally via polarimetry 
\citep{2013A&A...557A..82B,2012JPhCS.372a2073V}, although we did not attempt this here.

 The evolution is similar in the two cases -- the material outside the bow shock spirals towards the SMBH because of the drag, 
while inside the bow shock the cloud elements continue to move through the circumstellar environment. Figure~\ref{img_bowshock_star_dust} 
shows a very narrow shape of the bow shock near the pericentre. At this stage, many particles are torn away from 
the star, which is enhanced by both stellar wind and the outflow from the centre (see subsection \ref{truncation_radius}). However, a fraction that occupies the bow-shock region continues to orbit. In the case of strong wind from the centre, the particles are dragged more efficiently, so that a long tail forms behind the star. As mentioned above, only a 
diminishing part of the initial material survives the repeating passages through pericentre. The cloud is progressively 
stretched and accreted, and its centre drifts away from the position of the embedded star.

We also plot the changing division between the predicted position of the star and the centre of mass of the 
cloud (Fig.~\ref{img_bowshock_star_dust}, bottom row). The mutual separation tends to grow with time 
(apart from fluctuations) till the pericentre passage, then decreases again as the fraction of the cloud 
unbound to the star sublimates or becomes accreted onto the black hole. This offset 
is expected to be detectable in the NIR band if sufficient resolution is reached to resolve the cloud structure.

\begin{figure*}[tbh]
\centering
  \includegraphics[width=0.49\textwidth]{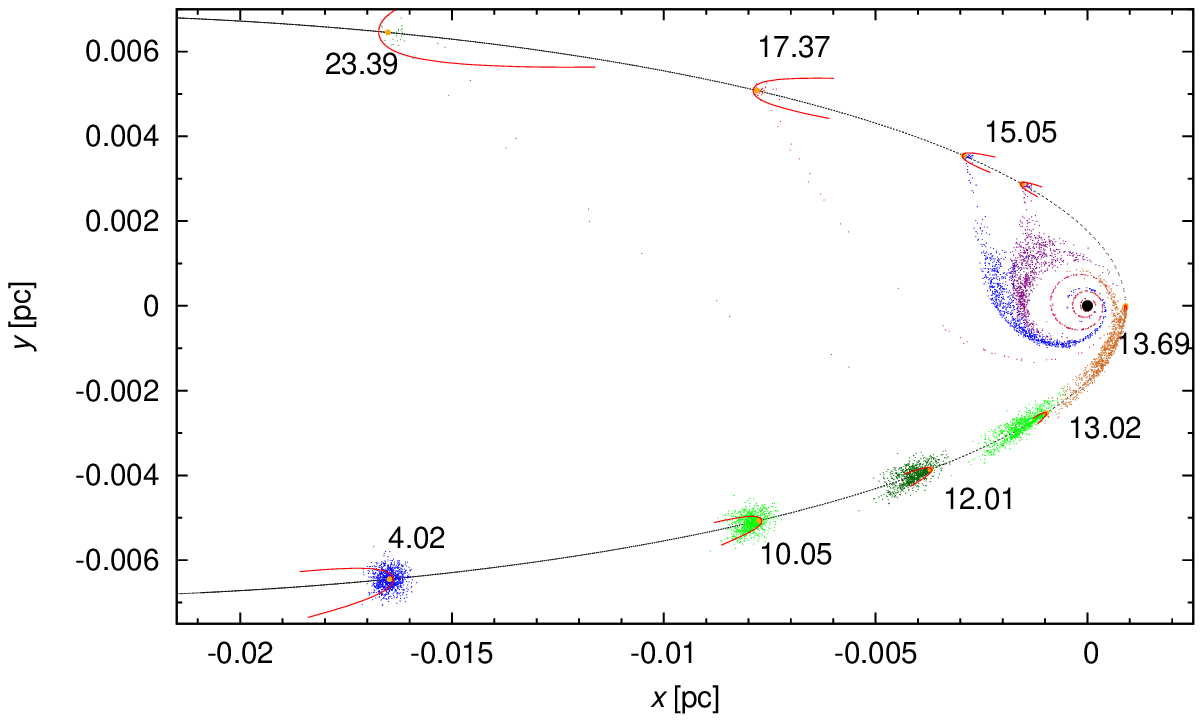} 
  \includegraphics[width=0.49\textwidth]{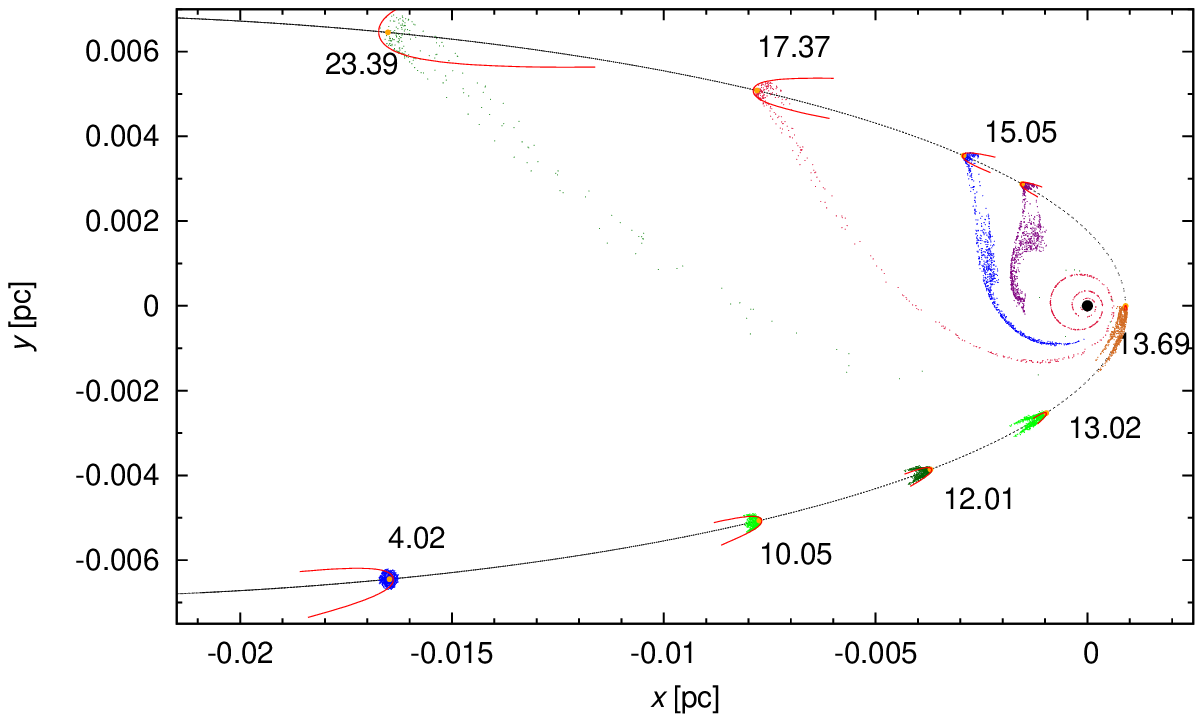} 
  \includegraphics[width=0.49\textwidth]{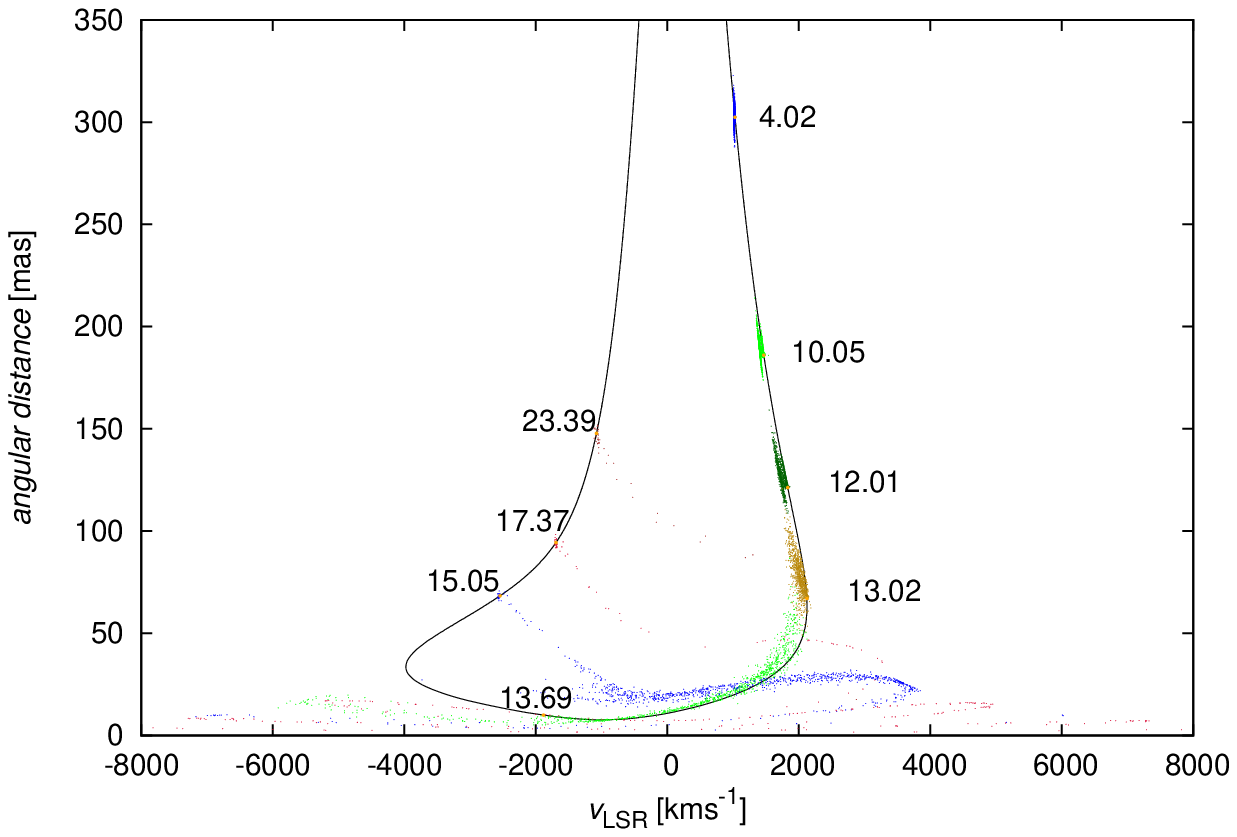}
  \includegraphics[width=0.49\textwidth]{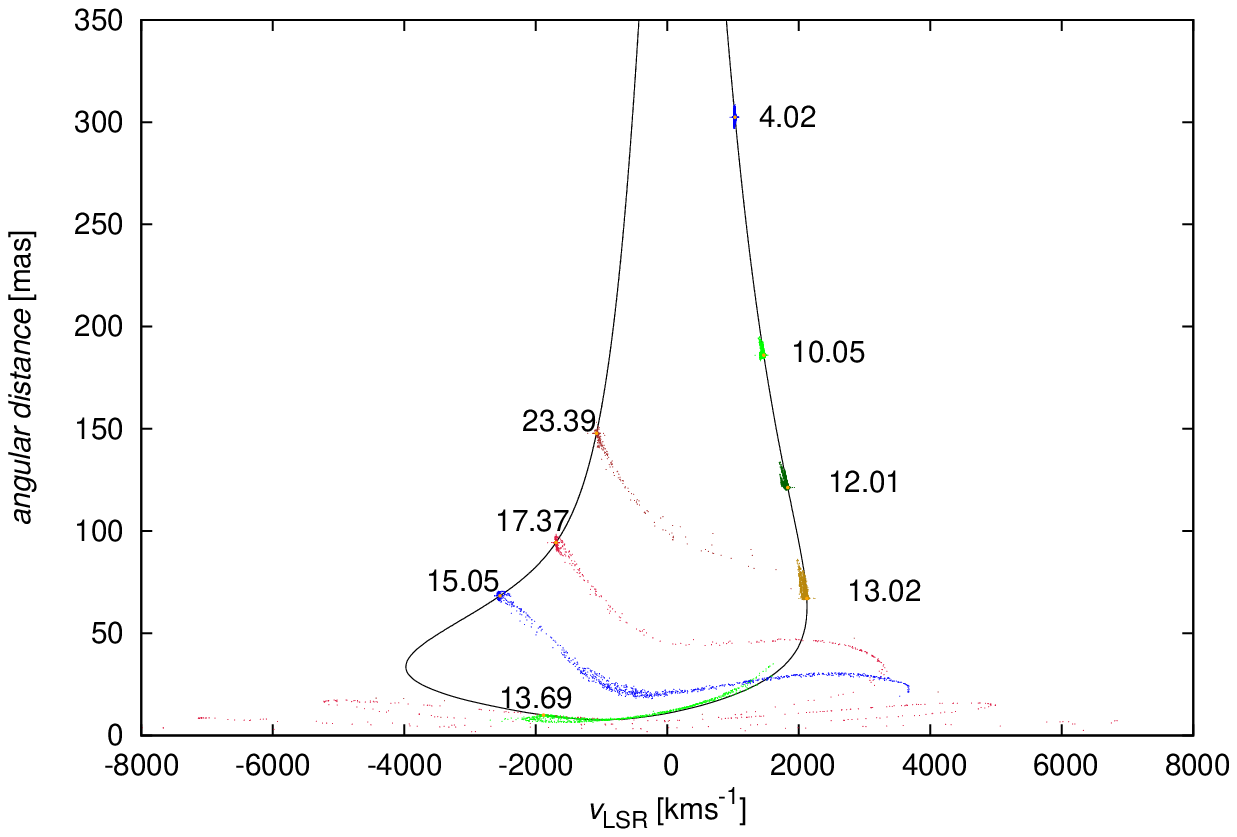} 
\caption{
Analogy to Fig. \ref{img_cloud510-6_distvel} for the case with the bow-shock.
\textit{Top left:} the evolution of an initially spherical Gaussian envelope in the presence of a star for chosen epochs since the start of the simulation. 
\textit{Bottom left:} representation of the evolution from the top--left panel is shown in the velocity--distance plane. 
The angular distance is expressed in miliarcseconds. 
\textit{Top right:} as in the top-left panel, but for a star-disc system. 
\textit{Bottom right:} the track in the velocity--position plane corresponds to the trajectory from the top--right panel.  
}
\label{img_star_disk_sphere_distvel}
\end{figure*}

\subsection{Spherical cloud vs. disc-like structure}
Near a protostar or T Tauri star, two types of dust/gas environment can be considered: a spherical cloud with an approximately 
Gaussian distribution in the position--velocity phase space, and a disc-like configuration with a Keplerian 
distribution of bulk velocity. The latter case represents a protoplanetary or a debris disc. \citet{2012NatCo...3E1049M} 
proposed such a disc as the origin of the cloud near SMBH. We compared the evolution of the two geometries.

 In both cases, dust grains are distributed around a $2{M_{\odot}}$ star with the parameters from the previous section. 
For the Gaussian velocity dispersion cloud, the initial FWHM was taken to be equal to $100\,\rm{AU}$
 ($\approx 12\,\rm{mas}$) and $5\,\rm{km/s}$,
respectively. Particles in the disc were distributed uniformly, with semi-major axes in the range $(0.1,50)\,\rm{AU}$, 
low eccentricity in the range $(0,0.1)$, and inclinations in the range $(0,30)^{\circ}$ and $(150,180)^{\circ}$ (taking into account both direct and retrograde orbits with respect to the orbit of the star).  

In figure~\ref{img_star_disk_sphere_distvel} we plot a typical evolution in the orbital plane for the chosen epochs (top panels) 
as well as the distribution in the position--velocity diagrams (bottom panels). These simulations indicate that a Gaussian 
cloud is not as much affected by the star as the particles in the disc. It is also evident that in case of the disc structure, 
more particles survive the pericentre passage and continue to
orbit the star. 

\begin{table*}[tbp]
\caption{Typical examples from different runs for the fraction of material deflected from the original trajectory during the first and second 
pericentre passages (evaluated with respect to the total number of particles in the cloud at the corresponding stage). 
Columns of model characteristics specify the initial distribution of particle positions and velocities (disc-like Keplerian vs.\
spherical Gaussian), presence or absence of the 
bow-shock effect and wind,  assumed mass of the star in the cloud core, and size of the dust grains.}
\centering
\begin{tabular}{cccccccc}
\hline\hline
\rule[-0.8em]{0pt}{2em} Run & \multicolumn{4}{c}{Model characteristics} & Note & \multicolumn{2}{c}{Mass capture $[\%]$} \\
\cmidrule(lr){2-5}\cmidrule(lr){7-8}
\rule[-0.8em]{0pt}{1em} & Initial distribution & Bow shock & $M_\star/M_{\odot}$ & Grain size $[\rm{\mu m}]$ & &1st passage & 2nd passage  \\
\hline 
\rule[-0.4em]{0pt}{1.5em}%
1 & disc-like (dr) & no & $3$ & $2.5$ & -- & $99.2$ & $0.0$ \\ 
2 & spherical & no & $3$ & $2.5$ & -- & $100.0$ & 0.0 \\ 
3 & disc-like (dr) & no & $3$ & $2.5$ & -- & $99.4$ & 0.0 \\ 
4 & disc-like (dr) & no & $3$ & $5$ & -- & $99.6$ & 0.0 \\ 
5 & disc-like (dr) & no & $2$ & $5$  & -- & $99.6$ & $50.0$ \\ 
6 & disc-like (dr) & no & $2$ & $5$ & -- & $99.5$ & $0.0$ \\ 
7 & disc-like (dr) & no & $3$ & $2.5$ & (i) & $95.2$ & $12.5$ \\ 
8 & disc-like (dr) & no & $3$ & $2.5$ & (i), (ii) 1000 km/s & $94.4$ & $14.3$ \\ 
9 & disc-like (dr) & yes & $2$ & $2.5$ & (ii) 500 km/s & 86.6 & $83.6$ \\ 
10 & disc-like (dr) & yes & $2$ & $2.5$ & (ii) 1000 km/s & $88.6$ & $58.8$ \\ 
11 & spherical & yes & $2$ & $2.5$ & (ii) 500 km/s & $99.2$ & $100.0$ \\ 
12 & disc-like (dr) & yes & $2$ & $2.5$ & (ii) 500 km/s & $91.3$ & $65.5$ \\ 
13 & disc-like (dr) & yes & $2$ & $0.6$ & (ii) 500 km/s & $91.3$ & $71.3$ \\ 
14 & disc-like (dr) & yes & $2$ & $0.1$ & (ii) 500 km/s & $98.1$ & $100.0$ \\ 
15 & disc-like (dr) & yes & $2$ & $2.5$ & (ii) 500 km/s & $87.3$ & $90.6$ \\ 
16 & sphere+disc (dr) & yes & $3$ & $2.5 $  & (ii), 1000 km/s, rad    & $93.3$ & $68.7$ \\
17 & disc-like (d)   & yes & $2$   & $2.5$  & (ii), 1000 km/s, rad   & $88.6$ & $74.8$  \\   
18 & disc-like (r)   & yes & $2$  & $2.5$   & (ii), 1000 km/s, rad   & $90.6$ &  $89.8$  \\ 
\hline
\end{tabular}
\newline
{\rule[-0.8em]{0pt}{2em}Notes: (i) dust region extending from $r=0.001\,\rm{AU}$;
(ii) central outflow wind $v_{\rm w}$ included; ``rad'' denotes radiation pressure from the star;\hfill~ \\[-4pt]
(d,r,dr) denotes only direct orbits, only retrograde orbits, and both, respectively.\hfill~ \\ 
Additional (non-essential) parameter is the number of numerical particles in each run (typically of the order of $10^3$). \hfill ~\\~}
\label{tab_captured_mass}
\end{table*}

\subsection{Fraction of mass influenced at subsequent encounters}
High-eccentricity passages are more likely to be non-repeating events if the cloud does not host a body inside its volume
(see Fig. \ref{img_cloud510-6_distvel}). The material of the cloud becomes dispersed and largely accreted onto the 
SMBH at the first pericentre passage. However, in the scenario with a shell surrounding a stellar object, a fraction of material 
remains bound to the star while only the rest is rerouted towards the black hole. In Table \ref{tab_captured_mass}, we summarise 
the typical results of computations of the percentage of the captured material for the first and the subsequent 
pericentre passages. 

\begin{figure*}[tbh]
\centering
\framebox{\includegraphics[width=0.28\textwidth]{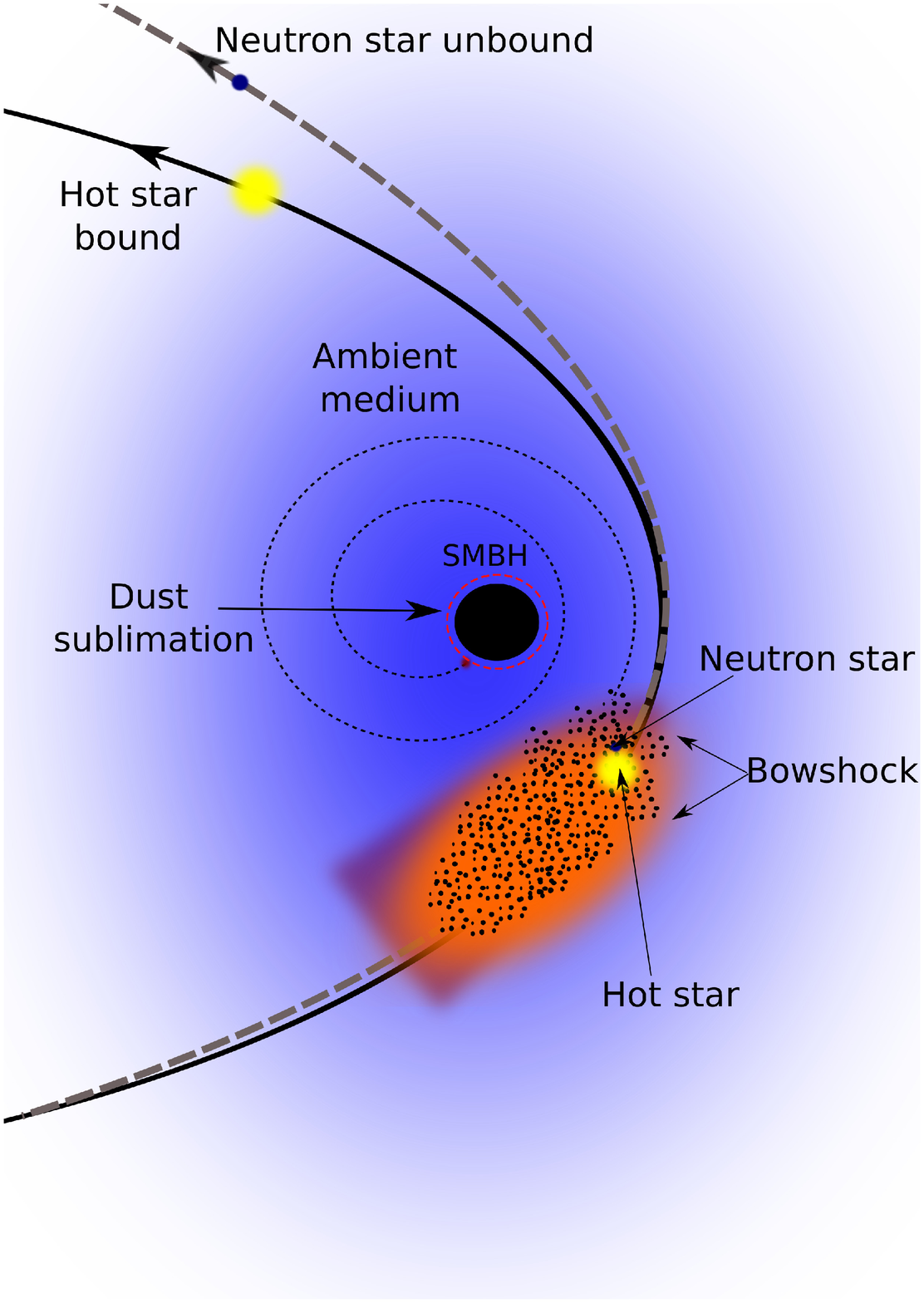}}
\hspace*{8mm}
\includegraphics[width=0.6\textwidth]{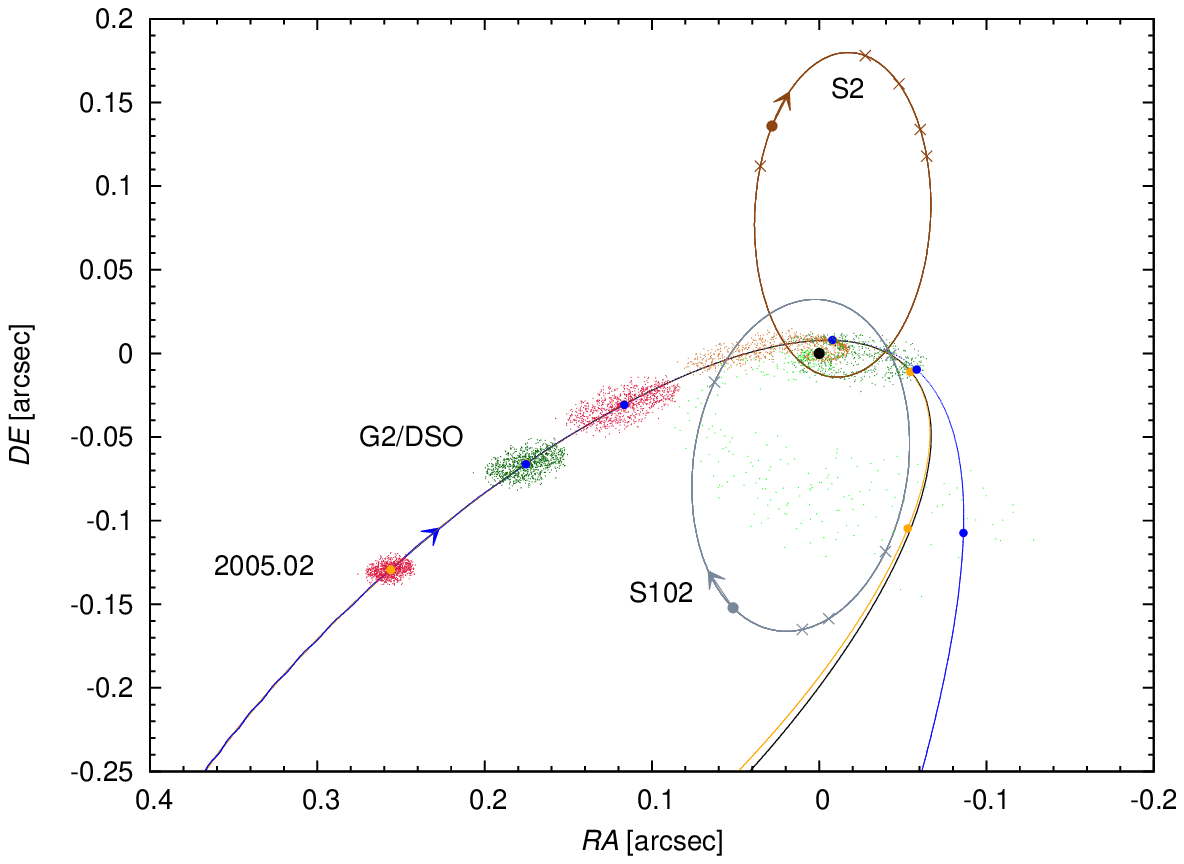}
\caption{Sketch of the model geometry for a binary with a common envelope (in the left panel)
and the orientation consistent with the G2/DSO nominal trajectory on the sky (in the right panel). The two components
(i.e., hot star+neutron star in this example) deviate from each other, the first one being
bound and the other one unbound at the moment of pericentre passage. The cloud
is then dispersed in the ambient medium and eventually accreted onto the SMBH.}
\label{illustration_binary}
\end{figure*}

As soon as the star/cloud on its trajectory reaches the critical radius for the mass overflow, the percentage of 
captured mass is high ($\gtrsim 90\%$ at the first pericentre passage). For the Gaussian distribution in the 
position--velocity phase space, it is near or equal to $100\%$ in both the bow-shock and no-bow-shock cases. When the bow shock is included and 
particles follow the bulk Keplerian distribution in a disc, more particles remain bound after the first pericentre crossing 
than the no-bow-shock scheme predicts (the bow-shock region protects some particles 
from being diverted from the cloud orbit near the pericentre). This is connected with the character of the drag forces on both sides
of the bow shock in our model and also agrees with previous results concerning the fate of a pressure
supported gas cloud, which appears to be particularly prone to complete destruction near the pericentre. 
However, one should bear in mind that we work with a four-parametric 
model. At the second pericentre passage, the total amount of the captured material is generally much smaller than at the first passage, nevertheless, the percentage fluctuates significantly in terms of the immediate mass of the cloud.

 We also tested a potential difference between the stability of direct (prograde) and retrograde discs. From the classical three-body theory, one can derive the difference between the critical Hill radii of direct and retrograde orbits, $r_{\rm{d}}$ and $r_{\rm{r}}$, respectively, resulting from the Coriolis term \citep{1979AJ.....84..960I}:
\begin{equation}
\frac{r_{\rm{r}}}{r_{\rm{d}}}=\left[\frac{5+e+2(4+e)^{1/2}}{3+e}\right]^{2/3}\,
\label{eq_hill_retrograde}
,\end{equation}
which leads to the ratio of $r_{\rm{r}}/r_{\rm{p}}\approx 1.9 $ for the eccentricity $e\approx 1$, which is our case. Hence, retrograde discs may not be as truncated as direct discs; the upper limit to the prolongation of their critical radius is the factor of $\sim 1.9$. This is consistent with our test runs involving only gravity. A direct disc around a $2$-$M_{\odot}$ star with the semi-major axes in the range $(0.05,10)\,\rm{AU}$, low inclinations $(0,10)^{\circ}$, and low eccentricities $(0,0.01)$ dissolves after one revolution after the first pericentre passage, forming families of orbits along the original trajectory and trajectories with smaller semi-major axes, with some particles escaping the system. A few of them remain in orbit of the star inside the Hill radius. A retrograde disc with the same distribution of orbital elements remains more compact and detached particles spread mostly along the original trajectory. However, when drag forces from both the stellar wind and the central outflow are involved, there is no difference in terms of stability between direct and retrograde discs (see runs 17 and 18 in Table \ref{tab_captured_mass}). This is because the stability of discs is no longer determined by the gravitational Hill radius, but by the smaller wind-truncation radius (see subsection \ref{truncation_radius}).

These results demonstrate that in the most cases the stellar envelope is significantly
affected around the pericentre, and the diverted part eventually becomes largely dispersed during the accretion event. The outcome is similar to the
case of a gaseous cloud, treated in the hydrodynamical regime \citep{2012ApJ...759..132A,2012ApJ...750...58B},
although a small fraction of the dusty envelope can survive to the following revolution.

\begin{table}[tbh]
\caption{Two exemplary cases of an embedded binary evolution. The true anomaly is the initial value, 
$\dot{m}_{\rm{w}}$ stands for the mass-loss rate. Additional wind parameters $v_{c}$ and $v_{\rm{w}}^{\star}$ denote the terminal 
velocities of the assumed spherical wind from the centre and the star, respectively.}
{\centering
\hfill\begin{tabular}{c|cc}
\hline
\hline
\rule[-0.4em]{0pt}{1.8em} Parameter & Case A & Case B \\
\hline 
\multirow{2}{*}{Mass of the binary components $[M_{\odot}]$} & \rule[-0.4em]{0pt}{1.5em} 4.0  & 3.0  \\
\rule[-0.7em]{0pt}{1.5em}            &            1.4  & 1.4  \\    
True anomaly $[^{\circ}]$ & 120.0 & 80.0 \\
Semi-major axis $[\rm{AU}]$ & 3.0  & 2.0  \\ 
Eccentricity & 0.05 & 0.05  \\ 
$\dot{m}_{\rm{w}}$ $[M_{\odot}\,\rm{yr^{-1}}]$ & \multicolumn{2}{c}{$10^{-7}$} \\
$v_{c}$ $[\rm{km\,s^{-1}}]$ & \multicolumn{2}{c}{500} \\
$v_{\rm{w}}^{\star}$ $[\rm{km\,s^{-1}}]$ & \multicolumn{2}{c}{700} \\
\hline 
\end{tabular}\hfill~}\\
{\rule[-0.2em]{0pt}{1.5em}Note:
In these examples we performed integration runs with $1000$ numerical particles representing the material of an initially spherical cloud,
whereas the circumbinary disc population consisted of $500$ particles with semi-major axes uniformly distributed from 
$10\,\rm{AU}$ to $50\,\rm{AU}$, inclinations up to $30^{\circ}$, and eccentricities ranging from $0.0$ to $0.1$. The spherical 
cloud population adopts a Gaussian distribution in the phase space with the initial FWHMs of $12.5\,\rm{mas}$ and $5\,\rm{km/s}$.
In both examples, the integration starts at the true anomaly of $-167^{\circ}$ for the barycentre position.\hfill}
\label{table_binary_param}
\end{table}

\subsection{Binary embedded within common envelope}

 The high eccentricity of the G2/DSO trajectory suggests that
it might be possible to connect this object with hypothetical events
of the three-body interaction \citep{1988Natur.331..687H} involving the SMBH
as the origin of stars on bound orbits near SMBH (\citeauthor{2003ApJ...592..935G}, \citeyear{2003ApJ...592..935G}, see figure~\ref{illustration_binary}). In this tentative scenario a binary system is disrupted during the 
pericentre passage. As a result of this interaction, one of the remnant 
bodies remains on a circularised orbit, whereas the other component gains orbital energy and is ejected 
\citep{2008ApJ...683L.151L,2012ApJ...749L..42B}. For realistic estimates of disruption rates, see \citet{2007ApJ...656..709P}.

Might the infalling cloud contain such a binary stellar system? The initial eccentric trajectory of the binary centre of 
mass was set in agreement with the infalling cloud within which it remain embedded. The orbital parameters of 
the binary system are expected to be disturbed at the moment of close encounter with the SMBH and the observational resolution about the nature
of the object should emerge soon after the pericentre passage. We can illustrate two qualitatively different options 
for the possible outcome of the post-pericentre evolution: (i) both components remain bound to the SMBH; (ii) one component 
is ejected from the system on a hyperbolic trajectory at the
expense of the orbital energy of the other component.

For each case we performed a number of simulations with a different setup of free parameters, namely, the masses $M_\star^{(i)}$ of 
the two components and the osculating elements of the embedded binary system. The primary component $M_{\star}^{(1)}$ represents 
a hot, wind-blowing star, the secondary $M_{\star}^{(2)}$ is thought to be a neutron star (for definiteness of the simulation).
The gaseous-dusty envelope enshrouding 
the whole binary was modelled in the same way as in previous examples, that is, in terms of dust particles that experience 
the hot atmosphere of the SMBH, the stellar wind from the primary and the bow-shock effect. The initial conditions and 
parameters are summarised in Table \ref{table_binary_param}. In general, the fraction of cloud initial mass that is retained
after the pericentre passage is diminished by the presence of the binary star in the core compared with an otherwise similar set-up
with a single stellar object embedded inside.

\begin{figure*}[tbh]
\centering
\begin{tabular}{cc}
\includegraphics[scale=0.6]{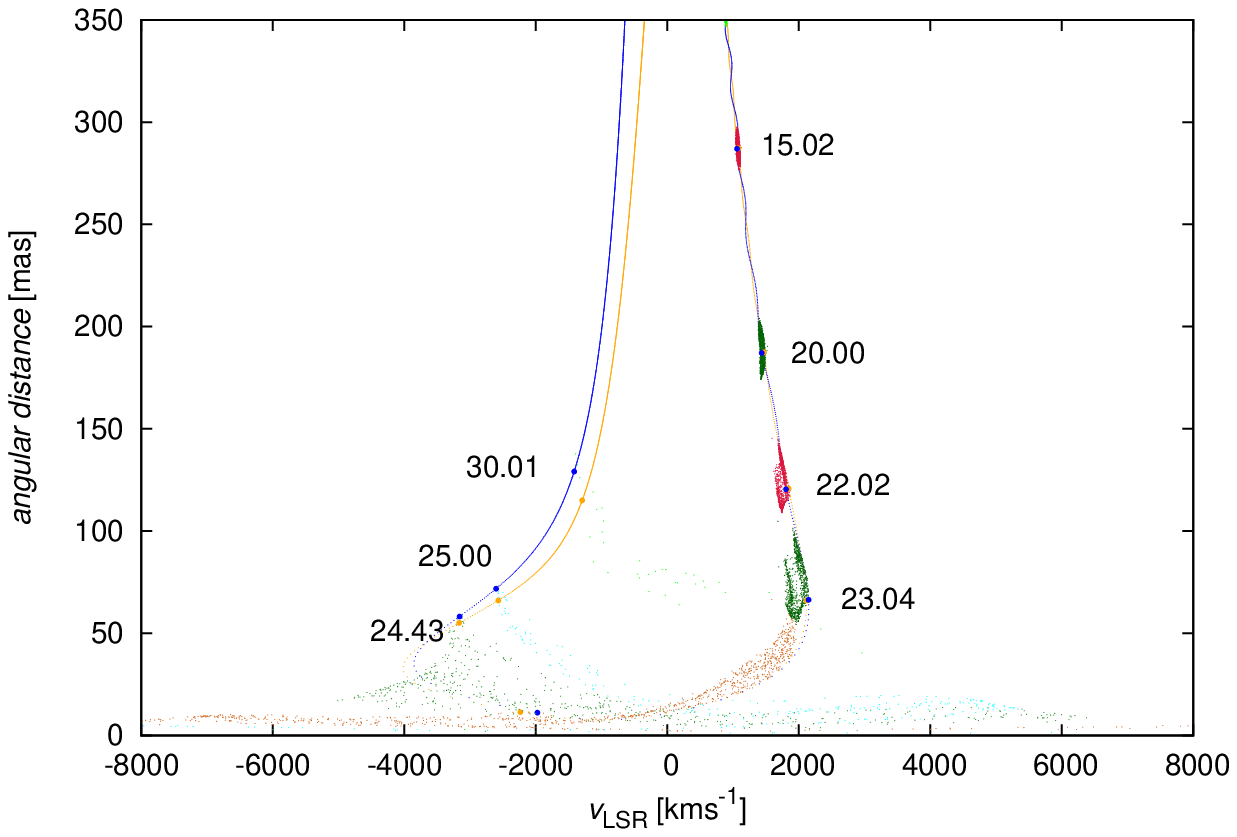} & \includegraphics[scale=0.6]{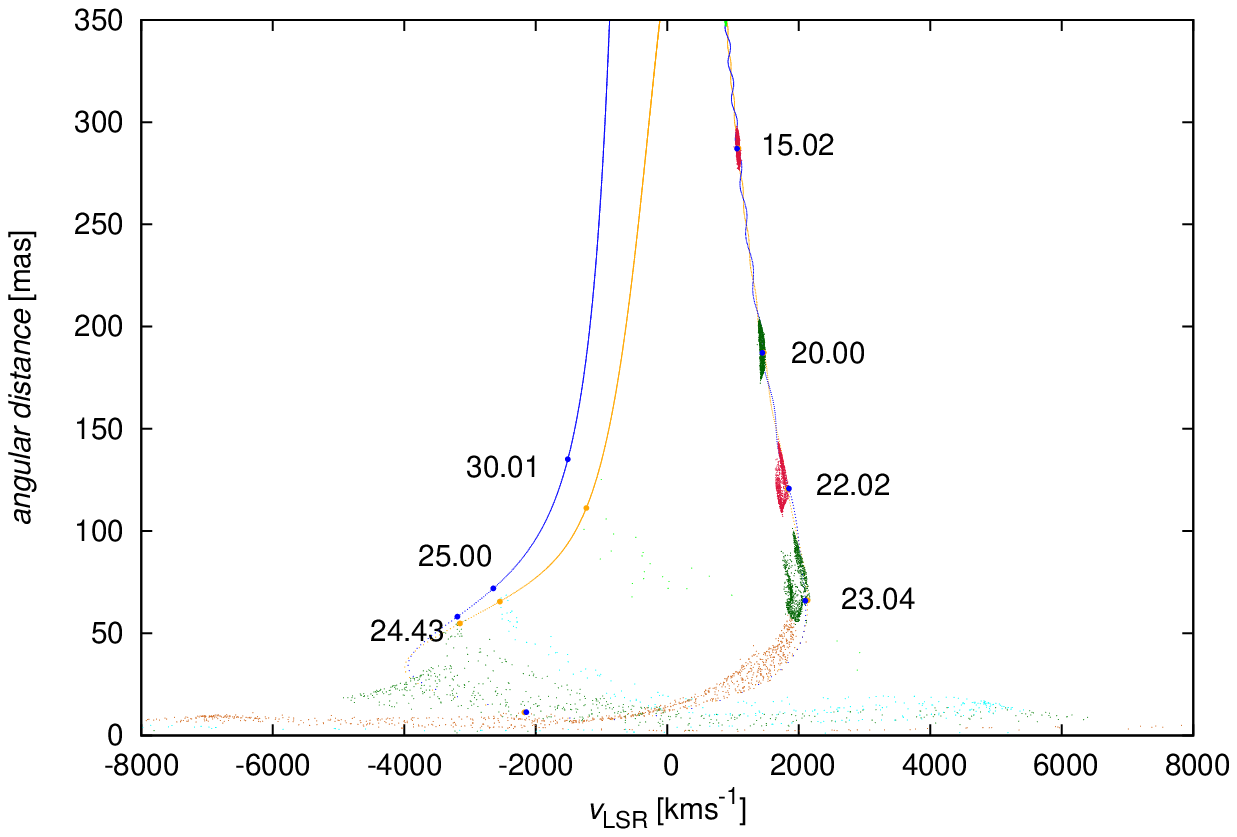}\\
\includegraphics[scale=0.65]{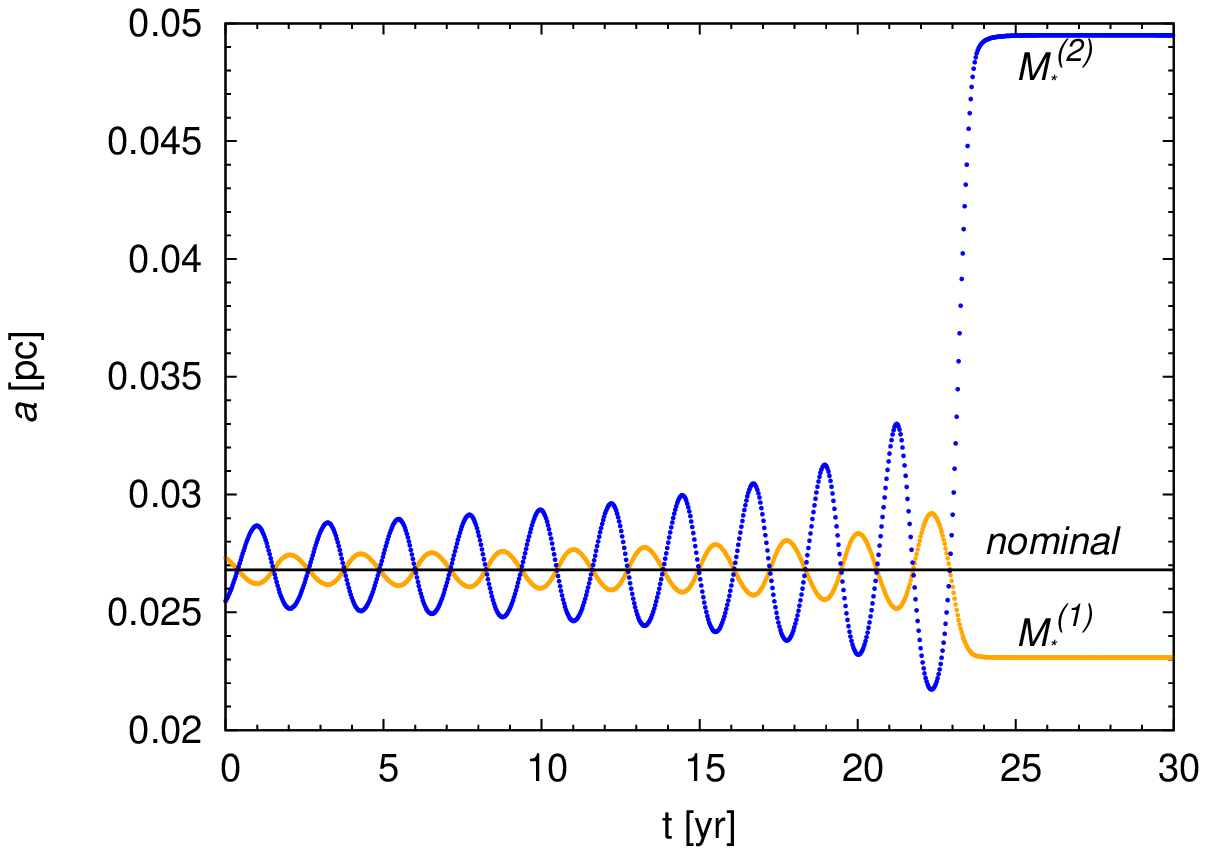} & \includegraphics[scale=0.65]{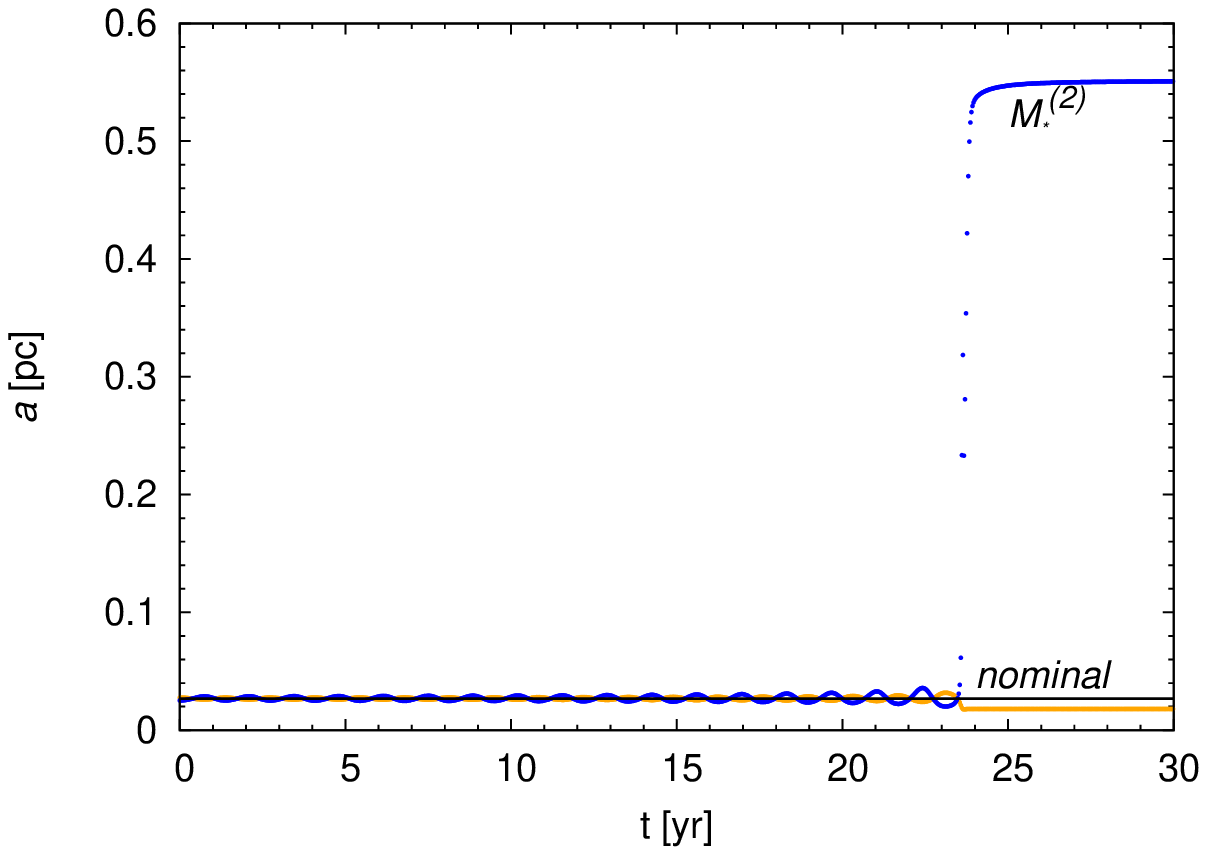}\\
\includegraphics[scale=0.65]{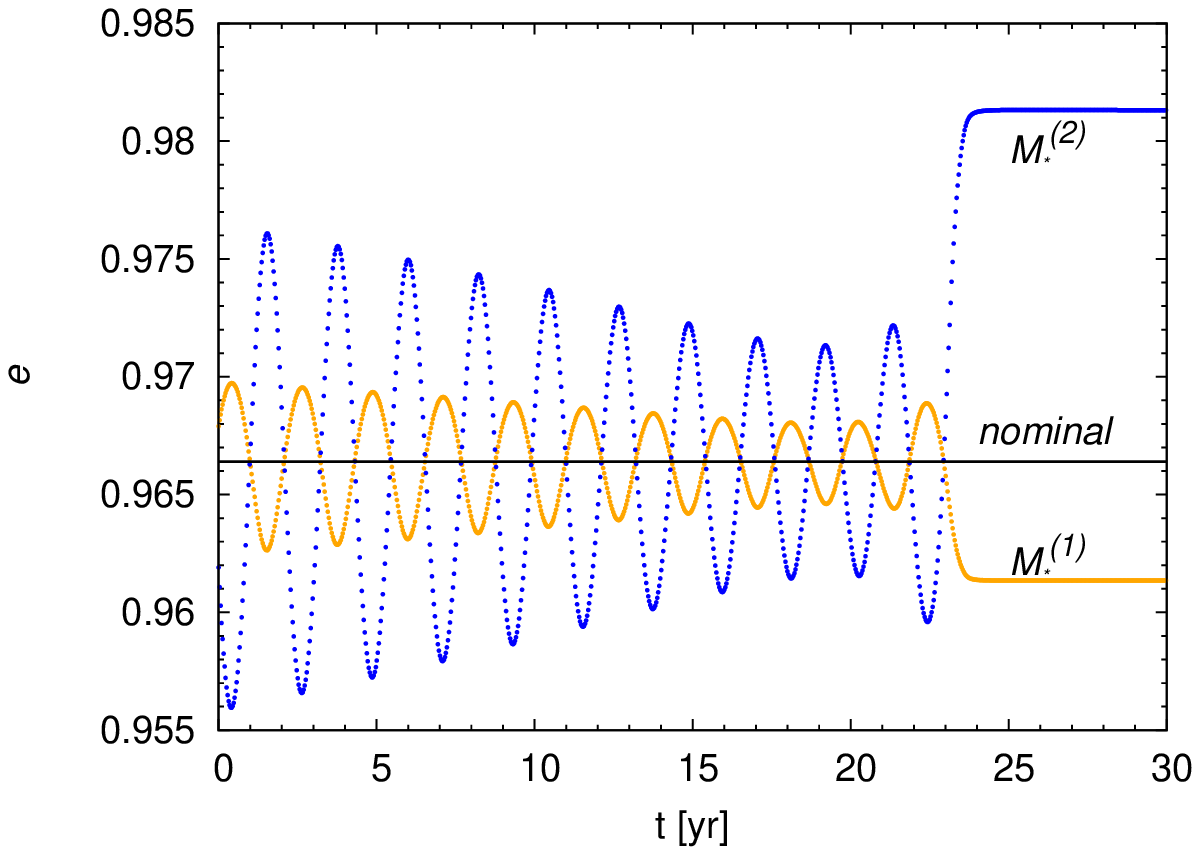} & \includegraphics[scale=0.65]{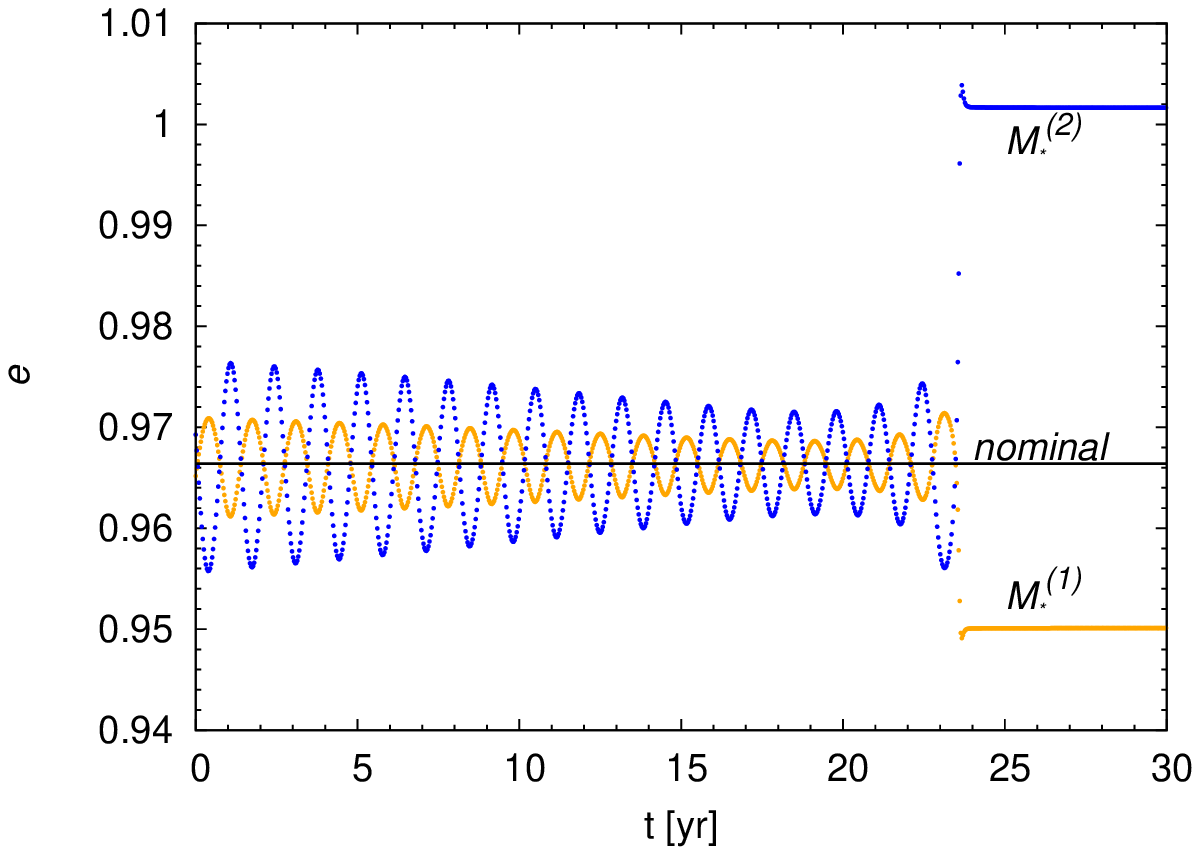}\\
\end{tabular}
\caption{Evolution of the cloud with an embedded binary, components of which become unbound via the three-body
interaction at the pericentre passage near the central SMBH. The initial trajectory of the binary barycentre corresponds
to the nominal trajectory of the G2/DSO (black solid line), whereas the post-pericentre stage depends strongly on the path of the 
individual components emerging from the system. Two exemplary cases are shown in different columns, 
corresponding to the choice of the component masses $M_{\star}^{(i)}$. \textit{Left panel:} $M_{\star}^{(1)}=4.0M_{\odot}$, 
$M_{\star}^{(2)}=1.4M_{\odot}$. \textit{Right panel:} $M_{\star}^{(1)}=3.0M_{\odot}$, $M_{\star}^{(2)}=1.4M_{\odot}$.}
\label{fig_bin_orbit}
\end{figure*}

In figure \ref{fig_bin_orbit} we show the distance--velocity plots, the temporal evolution of the binary semimajor axis $a$, and eccentricity $e$. 
At the pre-pericentre phase, the components orbit the common centre of mass, while at the pericentre they become 
unbound with respect to each other and start to move independently with different semi-major axes and eccentricities. We note,
however, that this is a multi-parametric system where the outcome of the evolution depends on a mutual interplay of different 
parameters.

The disruption event occurs when the secondary is outside the tidal-shearing radius, which may be again estimated using the Hill radius of the primary, eq. \eqref{eq_hill_rad}. For cases A and B, we derived $r_{\rm{H}} \sim 1.3\,\rm{AU}$ and $\sim 1.2\,\rm{AU}$, respectively. 
Following the discussion in the previous subsection, the disruption radius for the binary orbiting the black hole approximated by eq. \eqref{eq_hill_rad} is effectively enlarged if the secondary orbits the primary in a retrograde sense.

 The observation of a single event of the binary disruption while passing close to the SMBH is improbable and should be regarded only as a speculative scenario. However, other proposed scenarios for the G2/DSO infrared source, such as isothermal clouds or the disruption of a star, are also rare given the rate at which streams and winds collide or stars disrupt to produce such a cloud. Moreover, the nature of the binary content in the Galactic
centre is still unclear \citep{2012ApJ...757...27A}. Hence, all the estimates of binary replenishment in the central region are crude. The binary scenario can be easily rejected/confirmed by observations based on the post-pericentre evolution of the trajectory of the G2/DSO infrared source.

\section{Discussion}
\label{discussion}
We examined the pericentre passages and gradual destruction of dusty envelopes of stars that move supersonically through the immediate vicinity of the SMBH. Motivation for this topic arises because stars are shown to interact with the ionised medium close to Sgr A*, forming bow-shock structures (e.g., \citeauthor{2010A&A...521A..13M}, \citeyear{2010A&A...521A..13M}, and references therein). Here we focused on stars with a significant dust content that interacts with an optically thin wind outflow. Examples of such stars may also be found in the disc population of the Galaxy, specifically Herbig Ae/Be-type stars, see for instance \citeauthor{2009A&A...502..175B} (\citeyear{2009A&A...502..175B}).

The potential importance of the subject is heightened by the current passage of the infrared-excess G2/DSO source near Sgr~A*.  We focused on stellar-origin scenarios in which a star is enshrouded by a dusty shell. We revisited a simplified core-less scenario for comparisons of time scales and position-velocity distributions, although we did not perform detailed hydrodynamical simulations. The adopted approach is in several aspects complementary to hydrodynamical and MHD schemes, and it appears relevant for exploration of the dust component.
We computed the evolution in the presence of the star and the SMBH and also included additional effects within
an approximation:
  \begin{itemize}
  \item{hydrodynamical drag due to the plasma environment in the SMBH inner accretion zone;}
  \item{radially directed wind outflowing from the centre;}
  \item{wind-shearing in the immediate circumstellar environment;} 
  \item{bow-shock formation due to supersonic motion through the ISM.} 
  \end{itemize}   
We plotted the results showing the shape of a Keplerian disc-like system and a Gaussian envelope at different epochs. We 
computed the line-of-sight velocities (transformed to the local standard of rest) and the angular distance--velocity plots.

The idea of stars embedded in clouds of gas and dust does not have to be limited to our Galaxy. It has been explored
by various authors \citep[e.g.,][and references therein]{1994ApJ...434...46Z,1996ApJ...470..237A,1999A&A...352..452S,2002A&A...387..804V} 
in the context of repetitive interactions of stars of the nuclear cluster with the material of the accretion disc or a dusty torus in AGN.

The mini-spiral of the Galactic centre is a potential source of matter for infalling clouds. This structure contains a mixture of hot, warm, 
and cold phases. Under such conditions the accreted medium consists of complex plasma with a non-negligible content of dust 
\citep{1997ApJ...483..798C}. Located at a distance about $0.1$--$0.2$~pc (projected distance 
$\sim 0.06$~pc) from the supermassive black hole, this feature can be understood as consisting of three independent clumpy streams 
of mainly gaseous material at roughly Keplerian motion around the centre \citep{2010ApJ...723.1097Z,2012A&A...538A.127K}. The streams 
collide, and their mutual interaction may cause the loss of angular momentum and an occasional inflow of the clumpy material towards the 
black hole \citep{2005gbha.conf..197P,2013IAUS..290..199C}. In this scenario, the distribution of angular momentum determines the probability 
of setting the clouds on a plunging orbit \citep{2013A&A...555A..97C}.

One can thus expect that the current example of G2/DSO in the centre of the Milky Way may be a signature of a common 
mechanism that transports the gaseous clumpy medium and stars to the immediate vicinity of the central SMBH on sub-parsec scales.

\section{Conclusions}
\label{conclusions}
We modelled the fate of an infalling star with an extended envelope near the SMBH. The complex medium was treated 
in terms of numerical particles with their mass and size as parameters, interacting with the ambient environment. The mass of dust 
particles is typically large enough so that the gravitational influence on the grains
needs to be taken into account close to the SMBH and the mass-losing star.

 We assumed an orbit pericentre of the order of $10^3r_{\rm{s}}$, so that the star itself was not tidally disrupted. However,
the surrounding cloud was affected very significantly (core-less clouds are influenced even more, and they are
basically destroyed on the first encounter with the black hole). We noticed a significant mass-loss from the cloud at the 
first pericentre passage in all considered cases ($\gtrsim 90\%$). 
During the second passage the mass loss fluctuates. In other words, if the star is enshrouded in a dusty shell 
before the first pericentre passage, it becomes stripped of most of the envelope, unless the material is continuously replenished.
If a binary star is embedded within the envelope, there is a chance that the two components separate during the pericentre passage,
revealing the nature of the cloud core.

During the pericentre passage, the centre of mass of the cloud separates from the stellar core inside the cloud,
but then it returns to the star as the unbound particles are destroyed by sublimation or become accreted onto the
black hole.
The presence of the bow shock around a star somewhat diminishes the amount of captured particles (Table \ref{tab_captured_mass}), and
a greater fraction of the cloud can survive to the following pericentre passage. On the other hand,
the presence of a binary tends to dissolve the cloud more efficiently at the moment of close encounter with the SMBH. 
The characteristics of motion across the bow shock depend strongly on parameters, mainly the mass-loss rate 
and the stellar-wind velocity, so the predictions are uncertain and the outcome of the simulations vary. 
In addition, there are still uncertainties in the density and temperature profiles of the flow near the Galactic centre. 
However, the comparison of our simulations with post-pericentre observations can help to set better constraints.

\begin{acknowledgements} 
We thank Miroslav Bro\v{z} for discussions and help with the \texttt{Swift} integration package. We are grateful to Hagai B. Perets and Ladislav \v{S}ubr for  critical, constructive comments. The research leading to these results has received funding from the student grant of the Charles University in Prague (GAUK 879113), 
and the collaboration project of the Czech Science Foundation and Deutsche Forschungsgemeinschaft (GACR-DFG 13-00070J). 
The Astronomical Institute has been operated under the program RVO:67985815 in the Czech Republic. 
\end{acknowledgements}

\bibliographystyle{aa} 


\end{document}